\documentclass[aps,prd,tightenlines,twocolumn,amsmath,amssymb]{revtex4}%
\usepackage{graphicx}
\usepackage{epsfig}
\usepackage{subfigure}
\usepackage{graphics,color}
\usepackage[none]{hyphenat}
\usepackage{hyperref}
\usepackage{amsbsy}
\usepackage{verbatim}
\usepackage{bm}
\newcommand{\be}{\begin{equation}}
\newcommand{\ee}{\end{equation}}
\newcommand{\bea}{\begin{eqnarray}}
\newcommand{\eea}{\end{eqnarray}}
\newcommand{\nn}{\nonumber}

\begin{document}
\title{In-medium effect on the thermodynamics and transport coefficients in van 
der Waals hadron resonance gas }
\author{He-Xia Zhang}

\affiliation{Key Laboratory of Quark \& Lepton Physics (MOE) and Institute of 
Particle Physics, Central China Normal University, Wuhan 430079, China}
 
 \author{Jin-Wen Kang}
\affiliation{Key Laboratory of Quark \& Lepton Physics (MOE) and Institute of 
Particle Physics, Central China Normal University, Wuhan 430079, China}
 
 \author{Ben-Wei Zhang}
\email{bwzhang@mail.ccnu.edu.cn}
\affiliation{Key Laboratory of Quark \& Lepton Physics (MOE) and Institute of  Particle Physics, Central China Normal University, Wuhan 430079, China}
\affiliation{Institute of Quantum Matter, South China Normal University, Guangzhou 510006, China}
 
\begin{abstract}
	An extension of the van der Waals hadron resonance gas (VDWHRG) 
	model which includes in-medium thermal modification of hadron masses, the thermal VDWHRG (TVDWHRG) model, is considered in this paper. Based on the 2+1 flavor 
	Polyakov Linear Sigma Model (PLSM) and the scaling mass rule of hadrons we 	obtain the temperature behavior of all hadron masses for different fixed baryon chemical potentials $\mu_{B}$. We calculate various thermodynamic	observables at $\mu_{B}=0$ GeV in TVDWHRG model. 
	An improved agreement with the lattice data by TVDWHRG model in the 
	crossover region ($T\sim 0.16-0.19$ GeV) is observed as compared to those 
	by VDWHRG and Ideal HRG (IHRG) models.
	We further discuss the effects of in-medium modification of hadron masses 
	and VDW interactions between (anti)baryons on the dimensionless transport coefficients, such as shear viscosity to entropy density ratio
	($\eta/s$), scaled thermal ($\lambda/T^{2}$) and electrical ($\sigma_{el}/T$) 
	conductivities in IHRG model at different $\mu_{B}$, by utilizing 	 
	quasi-particle kinetic theory with relaxation time approximation. We find in contrast to the IHRG model, the TVDWHRG model leads to a qualitatively and quantitatively different behavior of transport coefficients with $T$ and $\mu_{B}$.

\end{abstract}
\maketitle

\section{INTRODUCTION}
\label{sec:intro}

Strongly interacting matter created in ultra-relativistic heavy-ion  experiments at the Relativistic Heavy-Ion Collider (RHIC) of BNL, and the Large Hadron Collider (LHC) of CERN has attracted intense theoretical and  experimental investigations. The study of strongly interacting matter can give a deep understanding of Quantum Chromodynamics (QCD) phase diagram and equation of state (EOS) of hot and dense matter.
Lattice QCD simulations as a reliable tool to study QCD thermodynamics have demonstrated that at finite temperature and vanishing baryon chemical potential $\mu_{B}$ there exists a smooth crossover (phase transition from hadronic 
matter to a chirally symmetric Quark-Gluon Plasma (QGP)) ranging from 0.15 to 0.2 GeV~\cite{Cheng:2006qk,Aoki:2006br}. 
Ideal Hadron Resonance Gas (IHRG) model is a widely used statistical model  which provides a remarkable good description of the Lattice QCD data~\cite{Borsanyi:2013bia,lattice-u,lattice0} at low temperature ($T< 0.15$ GeV) and zero $\mu_{B}$.
 However, IHRG model fails to fit with the Lattice QCD 
data in the crossover region ($T=0.16\sim0.19$ GeV). So an extended IHRG model called VDWHRG model which includes both the long distance attractive and the short distance repulsive van der Waals (VDW) type interactions between (anti)baryons 
is implemented~\cite{Vovchenko:2017cbu,Vovchenko:2016rkn}. The results of thermodynamic quantities within VDWHRG model are closer to the Lattice QCD data in crossover region than that within IHRG model.

The transport properties of strongly interacting matter play a significant role in describing the dynamical evolution of hot and dense matter. Shear viscosity for hadronic sector has been 
analytically calculated in the relativistic kinetic theory using Chapman-Enskog (CE) approximation~\cite{shear-CE-Pion,shear-CE1,shear-CE2,st-Bol-CE,shear-CE3,shear-CE4} and relaxation time approximation (RTA)~\cite{RTA1,RTA2,RTA3,RTA4}. In Ref.~\cite{shear-liner response-pion}, shear viscosity for pion gas has been obtained in the linear response theory using Kubo formulas.
The shear viscosity for the hadronic phase has also been computed  in  microscopic transport model (e.g. SMASH~\cite{shear-SMASH}, UrQMD~\cite{shear-UrQMD}, B3D transport model~\cite{shear-B3D} and PHSD~\cite{shear-PHSD}), in exclude volume HRG (EVHRG) model~\cite{shear-evhrg1,shear-evhrg3,shear-evhrg4}, in the  chiral perturbative theory (ChPT)~\cite{shear-chpt,sek-cpt-meson,sek-cpt-pion1}, in effective QCD models ~\cite{est-NJL,sbet-njl,sbet-PLSM,sbe-PQM model},  in quasi-particle theory~\cite{shear-quasi1,shear-quasi2}, in Scaled Hadron Masses-Couplings (SHMC) model~\cite{SHMC}, and so on.
A few articles also deal with electrical conductivity in pure pion gas~\cite{electric-kinetic-pion,sek-cpt-pion1}. In  hadronic matter, electrical conductivity has been estimated by employing the relativistic kinetic theory ~\cite{electric-kinetic1,electric-kinetic3} and Kubo formalism~\cite{electric-kubo}.
Furthermore, electrical conductivity in  hadronic temperture domain also recently  computed in transport code SMASH~\cite{electric-smash}, in PHSD  transport model~\cite{electric-PHSD1,electric-PHSD2}, in anisotropic lattice QCD simulation~\cite{electric-lattice}. Another 
important but less concerned transport coefficient is
 thermal conductivitie  which have been calculated in  hot pion gas~\cite{RTA1,RTA2,thermal-kinetic-pion,sbt-pion-buu-CE,thermal-pion} and hadronic gas mixture~\cite{thermal-kinetic-kaon-phonon,thermal-kintic-hadronmix,st-Bol-CE}  using kinetic theory.  Recently, the electrical and thermal conductivities  of hadronic temperature domain  have also been estimated in effective QCD models~\cite{est-NJL,sbt-PQM model,Harutyunyan:2017ttz} and EVHRG model~\cite{electric and thermal evhrg}.
However, so far most of these  calculations have taken the vacuum hadron masses as  inputs and have not taken into account the influence of in-medium hadron masses on transport coefficients.

As we know that spontaneous chiral symmetry  breaking is an important feature in QCD vacuum, which is related to  the generation of hadron masses~\cite{0,1,2}.
With the increase of temperature or baryon chemical potential, chiral symmetry will be restored, which implies that the masses of constituent quarks should be reduced to be zero.
Once the constituent quark masses are relevant to temperature and baryon chemical potential, the masses of subsequent hadrons also should be dependent of temperature and baryon chemical potential naturally. 
In the literature two main effective QCD-like models, the Polyakov Nambu-Jona-Lasinio (PNJL) (e.g.~\cite{Costa:2003uu,NJLCosta:2008dp}) and Polyakov linear sigma model (PLSM) (e.g.~\cite{Tawfik:2019rdd,Tawfik:2014gga,Tawfik:2015tga,Tiwari:2013pg,Mao:2009aq})
are widely used.
These models are successful in explaining the dynamics of both chiral symmetry  breaking-restoration and the confinement-deconfinement transition, as well as can  describe the thermal evolution of meson masses in hot and dense QCD matter.
So it is of great interest to replace vacuum hadron masses with temperature and chemical potential dependent masses to explore thermal hadron mass effect on the thermodynamic quantities and transport coefficients in hot and dense hadronic 
matter.

In this work, we  develop a TVDWHRG model, which is an extension of VDWHRG model by including the dependence of hadron masses on temperature $T$ and baryon chemical potential $\mu_B$. In TVDWHRG model we utilize the 2+1 flavor Polyakov Linear Sigma Model (PLSM)  combined with the generalized mass scaling rule of hadrons to obtain the thermal behavior of hadron masses. Then these thermal hadron masses are taken as dynamic inputs to calculate the thermodynamic quantities in VDWHRG model. We further explore how the effects of thermal hadron masses and VDW interactions influence  the transport coefficients, such as  shear viscosity, electrical and thermal conductivities in  hadronic matter. In our model the derivation of transport coefficients is performed by solving the  Boltzmann equation in  relaxation time approximation.

The paper is organized as follows. In Sec.~\ref{sec:HRGM} we review the ideal and interacting HRG  models. In Sec.~\ref{sec:medium-modifi}, we give a brief overview on PLSM and discuss the analytical expressions for the medium modifications of hadron masses at finite temperature and baryon chemical potential.
 In Sec.~\ref{sec:transport-coefficients}, we present the formulas  of the transport coefficients in the quasi-particle kinetic theory under relaxation time approximation.
  In Sec.~\ref{sec:results} the numerical results and discussions are presented. And 
   Sec.~\ref{sec:conclusions} summarizes our studies.


\section{HADRON RESONANCE GAS }
\label{sec:HRGM}

\subsection{Ideal hadron resonance gas model }

In IHRG model all thermodynamic quantities  can be obtained from the sum of the logarithm of grand canonical partition function over all hadrons and resonances~\cite{Andronic:2012ut}
\begin{equation}
\ln Z^{id}=\sum_{i}^{}\ln Z_{i}^{id}(T,\mu_{i},m_{i}).
\end{equation}
For particle species $i$,
\begin{equation}
\ln Z_{i}^{id}=\pm\frac{Vg_{i}}{(2\pi)^{3}}\int d^{3}p\ln\left[1\pm 
\exp(-(E_{i}-\mu_{i})/T)\right] .
\end{equation}
Here $id$ refers to the ideal (non-interacting) gas and $V$ is the volume of  system, $g_{i}$ stands for the degeneracy factor which satisfies the relation $g_{i}=(2J_{i}+1)$, $J_{i}$ is angular momentum of hadron species $i$,
 the sign $\pm$ is positive for fermions and negative for bosons.
$E_{i}=\sqrt{p^{2}+m_{i}^{2}}$ denotes energy of the single particle. 
$m_{i}$ presents mass of hadron species $i$, which is usually taken as the vacuum hadron mass. In this paper, we also consider the effects of finite temperature and chemical potential on masses of hadrons.
$\mu_{i}=B_{i}\mu_{B}+S_{i}\mu_{S}+Q_{i}\mu_{Q}$ is the chemical potential of particle species $i$, where $B_{i},S_{i},Q_{i}$ are the baryon number, strangeness  and electric charge respectively, $\mu_{B/S/Q}$ gives the corresponding chemical potential. 
We assume $\mu_{S}=\mu_{Q}=0$, which is a reasonable approximation in heavy-ion collision experiments~\cite{assume}.
The thermodynamic quantities (the pressure, the energy density and the number density) in IHRG model can be given by~\cite{Sarkar:2018mbk,Mohapatra:2019mcl}
\begin{eqnarray}
P^{id} &=& T\frac{\partial \ln Z^{id}}{\partial V} = 
\sum_{i}^{}g_{i}\int\frac{d^{3}p}{(2\pi)^{3}}\frac{p^{2}}{3E_{i}}f_{i}^{id},\\
\epsilon^{id} &=&-\frac{1}{V}\left(\frac{\partial \ln Z^{id}} {\partial
\frac{1}{T}}\right)_{\frac{\mu_{i}}{T}}=\sum_{i}^{}g_{i}\int
\frac{d^{3}p}{(2\pi)^{3}}E_{i}f_{i}^{id},\\
n^{id} &=& \frac{T}{V}\left(\frac{\partial \ln Z^{id}}{\partial \mu_{i}}
\right)_{V,T}=\sum_{i}^{}g_{i}\int\frac{d^{3}p}{(2\pi)^{3}}f_{i}^{id},
\end{eqnarray}
where $f_{i}^{id}$ is ideal Fermi or Bose distribution function 
$f_{i}^{id}(T,p,\mu_{i})=1/(\exp[(E_{i}-\mu_{i})/T]\pm1)$.

\subsection{Interacting hadron resonance gas }

In this work, we also  consider a  more realistic system, where   the short-distance repulsive interaction and the long-distance attractive interaction exist among hadrons. 
There are different phenomenological excluded-volume models 
to simulate the repulsive interaction of hadrons such as van der 
Waals~\cite{Greiner1995} and Carnahan-Starling excluded-volume models~\cite{carnahan-starling} with the effect of quantum 
statistics. For the attractive interaction, four various forms have been discussed~\cite{Vovchenko:2016rkn,Samanta:2017yhh,peng,kwong}$:$ van der Waals, Redlich-Kwong-Soave, Peng-Robinson and Clausius models. Therefore, to take into account both the repulsive interaction and attractive interaction, eight interacting hadron resonance gas models could be employed: the VDW, RKS, PR, Clausius, 
VDW-CS, RKS-CS, PR-CS and Clausius-CS models.
In interacting hadron resonance gas model, the repulsive and attractive  interactions only exist between   baryon-baryon pairs and between antibaryon-antibaryon pairs while the baryon-antibaryon, meson-baryon and meson-meson  interactions are neglected~\cite{Vovchenko:2016rkn,Vovchenko:2017cbu}.
 So the total pressure in grand canonical ensemble can 
be written as~\cite{Vovchenko:2017cbu}
\begin{equation}
P(T,\mu)=P_{M}(T,\mu)+P_{B}(T,\mu)+P_{\bar{B}}(T,\mu),
\end{equation}
with
\begin{eqnarray}
P_{M}(T,\mu)&=&\sum_{z\in M}^{}P_{z}^{id}(T,\mu_{z}),\\
P_{B}(T,\mu)&=&[F(h_{B})-h_{B}F'(h_{B})]\sum_{z\in B}^{}
P_{z}^{id}(T,\mu_{z}^{B*})\nonumber\\
&&+n_{B}^{2}u'(n_{B}),\\
P_{\bar{B}}(T,\mu)&=&[F(h_{\bar{B}})-h_{\bar{B}}F'(h_{\bar{B}})]
\sum_{z\in\bar{B}}^{}P_{z}^{id}(T,\mu_{z}^{\bar{B}*})\nonumber\\
&&+n_{\bar{B}}^{2}u'(n_{\bar{B}}),
\end{eqnarray}
where $\mu$ is  baryon chemical potential in current work,  the subscripts $M$,~$B$,~$\bar{B}$ stand for mesons, baryons and antibaryons, respectively. The constructed functions $F(h_{B(\bar{B})})$ and 
$u(n_{B(\bar{B})})$ are related to the repulsive and attractive interactions between
(anti)baryon pairs, respectively. The analytical forms of 
$F(h_{B(\bar{B})})$ 
and $u(n_{B(\bar{B})})$ are different according to the choice of interacting hadron resonance gas models listed 
previously. $h_{B(\bar{B})}$ denotes the
packing ratio of all (anti)baryonic volume occupied in total system volume 
which satisfies the relation of $h_{B(\bar{B})}=\frac{b}{4}n_{B(\bar{B})}$. $n_{B(\bar{B})}$  is the total number  density of 
(anti)baryons,  which can be obtained by using 
$n_{B}=\partial P_{B}/\partial \mu_{z}$,
\begin{equation}\label{eq:n}
n_{B}(T,\mu)=F(h_{B})\sum_{z\in B}^{}n_{z}^{id}(T,\mu_{z}^{B*}).\\
\end{equation}
And the shifted chemical potential of baryon $\mu_{z}^{B*}$ is given 
as~\cite{Vovchenko:2017cbu} 
\begin{eqnarray}\label{eq:u}
\mu_{z}^{B*}-\mu_{z}=\frac{b}{4}F'(h_{B})\sum_{z\in 	
B}^{}P_{z}^{id}(T,\mu_{z}^{B*})\nonumber \\
-u(n_{B})-n_{B}u'(n_{B}).
\end{eqnarray}
The key is to obtain $\mu_{z}^{B*}$. At given $T$ and $\mu$,  $\mu_{z}^{B*}$ 
can be calculated by solving Eqs.~(\ref{eq:n})-(\ref{eq:u}) numerically. 
Accordingly, other thermodynamic quantities such as the entropy density 
$s_{B}=(\partial P_{B}/\partial T)_{\mu}$ and the energy density can be 
determined by 
\begin{eqnarray}\label{eq:sande}
s_{B}(T,\mu)&=&F(h_{B})\sum_{z\in B}^{}s^{id}_{z}(T,\mu_{z}^{B*}),
\end{eqnarray}
and
\begin{eqnarray}\label{eq:e}
\epsilon_{B}(T,\mu)&=&F(h_{B})\sum_{z\in B}^{}\epsilon^{id}_{z}(T,\mu_{z}^{B*})
+n_{B}u(n_{B}).
\end{eqnarray}
 Eqs.~(\ref{eq:n})-(\ref{eq:e}) are also applicable to antibaryons. In this work, 
we use VDW model in which $F(h_{B(\bar{B})})=1-4h_{B(\bar{B})}$ and 
$u(n_{B(\bar{B})})=-an_{B(\bar{B})}$.
The parameters $a$ and $b$ are determined by reproducing the properties of 
nuclear matter in  ground state~\cite{groundstate}, according to the choice of interacting hadron resonance gas models~\cite{VVgroundstate}.


\section{Mass sensitivity of hadrons at finite temperature and baryon chemical potential}
\label{sec:medium-modifi}

\subsection{The Polyakov linear sigma model (PLSM)}
As mentioned in Sec.~\ref{sec:intro}, the melting behavior of hadron masses is related to the temperature and chemical potential dependent  constituent quarks. In present work, the  dynamical information  of constituent quark masses  in QCD medium  can be determined by empolying the SU(3) Polyakov linear sigma model (PLSM).
We next briefly introduce  the linear sigma model with $N_{f}=2+1$ flavor quarks, coupled to the Polyakov loop dynamics to formulate the PLSM.  The related Lagrangian is given as~\cite{Tawfik:2019rdd,Tawfik:2014gga,Tawfik:2015tga,Tiwari:2013pg,Mao:2009aq},
\begin{equation}\label{plsm}
\mathcal{L}=\mathcal{L}_{chiral}-\mathcal{U}(\phi,\phi^*,T),
\end{equation}
where the chiral part of the Lagrangian, $\mathcal{L}_{\rm chiral}=\mathcal{L}_{quark}+\mathcal{L}_{meson}$, has $SU(3)_{L}\times SU(3)_{R}$ symmetry~(details can be found in~\cite{Schaefer:2008hk}). The first term in  $\mathcal{L}_{\rm chiral}$ corresponds to the fermionic contributions from quarks, and the second term represents the mesonic contribution, both contributions have been extensively discussed in Refs.~\cite{Tawfik:2014gga,Tawfik:2015tga,Tiwari:2013pg,Mao:2009aq}.
The second term in Eq.~(\ref{plsm}), $\mathbf{\mathcal{U}}(\phi, \phi^*, T)$, represents the Polyakov-loop effective potential to introduce gluon degrees of freedom and the dynamics of the quark-gluon interactions \cite{Polyakov:1978vu}, which is expressed by using the dynamics of the thermal expectation value of a color traced Wilson loop in the temporal direction. Correspondingly, the traced Polyakov-loop  variable and its conjugate can read
\begin{eqnarray}
\phi =\langle\mathrm{Tr}_c \,L\rangle/N_c, \label{phais1},\qquad
\phi^* = \langle\mathrm{Tr}_c\, L^{\dag}\rangle /N_c, \label{phais2}
\end{eqnarray}
where $L$ is the Polyakov loop.  And $L$ can be represented by a matrix in the color space \cite{Polyakov:1978vu}
\begin{eqnarray}
L(\vec{x})=\mathcal{P}\mathrm{exp}\left[i\int_0^{\beta}d \tau A_0(\vec{x}, \tau)\right],\label{loop}
\end{eqnarray}
where $\beta=1/T$ denotes the inverse temperature, $\mathcal{P}$  and  $A_0$ are path ordering and temporal component of Euclidean vector field, respectively \cite{Polyakov:1978vu}. At vanishing chemical potential, $\phi=\phi^{*}$, the Polyakov loop  is recognized as an  order parameter for the deconfinement phase-transition. In this work, we use a logarithmic formed Polyakov-loop effective potential~\cite{Ratti:2005jh}, which is motivated by the underlying QCD symmetries in the pure gauge limit.
\begin{eqnarray}
\frac{\mathcal{U}_{Log}(\phi,\phi^*,T)}{T^{4}}&=&\frac{-a(T)}{2}\phi^*\phi+b(T)\ln[1-6\phi^{*}\phi\nn\\
&&+4(\phi^{*3}+\phi^3)-3(\phi^*\phi)^2],
\end{eqnarray} 
with 
\begin{eqnarray}\label{eq: parameter}
a(T)&=&a_{0}+a_{1}(T_{0}/T)+a_{2}(T_{0}/T)^2, \nn\\ b(T)&=&b_{3}(T_{0}/T)^3.
\end{eqnarray}
In Eq.~(\ref{eq: parameter})  $a_{0}=3.51$, $a_{1}=-2.47$, $a_{2}=15.2$, $b_{3}=-1.75$,  which are determined by fitting pure gauge lattice  datas~\cite{Ratti:2005jh}.
 $T_{0}=270$~MeV is the critical temperature for the deconfinement in  Yang-Mills theory.
In the mean field approximation~\cite{Tawfik:2014gga},  the grand canonical potential of PLSM can be written as
\begin{eqnarray}
\Omega(T, \mu_{fl}) &=& U(\sigma_x, \sigma_y)+\mathbf{\mathcal{U}}(\phi, \phi^*, T) + \Omega_{\bar{q}q} (T,\mu_{fl},\phi,\phi^{*}), 
\label{potential}\nn\\
\end{eqnarray}
where $\sigma_x$ and $\sigma_y$ stand for the non-strange and strange chiral condensates. The first term in Eq.~(\ref{potential}), the purely mesonic potential, is given as
\begin{eqnarray}
U &=&-h_x
\sigma_x-h_y \sigma_y+ \frac{m^2(\sigma^2_x+\sigma^2_y)}{2} -\frac{c\sigma^2_x \sigma_y }{2\sqrt{2}} \nonumber \\
&&+ \frac{\lambda_1 \sigma^2_x \sigma^2_y}{2} +\frac{(2 \lambda_1
	+\lambda_2)\sigma^4_x}{8}  + \frac{ (\lambda_1+\lambda_2)\sigma^4_y}{4}.
\end{eqnarray}
Here, $m^2$, $h_x$, $h_y$, $\lambda_1$, $\lambda_2$ and $c$ are model parameters as reported in Ref.~\cite{Schaefer:2008hk}. The parameters  used in the present work are listed in Table.~\ref{par_tab1}. 
The third term in Eq.~(\ref{potential}), $ \Omega_{\bar{q}q}(T,\mu_{fl};\phi,\phi^{*}) $,  is the quark-antiquark potential, which  can be shown as ~\cite{Tawfik:2014gga}
\begin{eqnarray} \label{z_MF}
\Omega_{\bar{q}q} 
=-2 T \sum_{fl=u,d,s} \int \dfrac{d^3 p}{(2 \pi)^3} (\ln g_{fl}^{+}+\ln g_{fl}^{-}).
\end{eqnarray} 
The expressions of $g_{fl}^+$ and $g_{fl}^{-}$ are defined as
\begin{eqnarray}
g_{fl}^{+}=[ 1+3(\phi+\phi^* e^{-E_{fl}^+/T}) e^{-E_{fl}^+/T}+e^{-3 E_{fl}^+/T}], \\
g_{fl}^{-}=[ 1+3(\phi^*+\phi e^{-E_{fl}^-/T}) e^{-E_{fl}^-/T}+e^{-3 E_{fl}^-/T}],
\end{eqnarray}
where $E_{fl}^\pm=E_{fl}\mp\mu_{fl}$, $E_{fl}=\sqrt{p^2+m_{fl}^{2}}$  is the single particle energy   with  the flavor-dependent  constituent (anti)quark mass $m_{fl}$. For a symmetric quark matter, we take the uniform blind chemical potential, i.e. $\mu_{fl}\equiv\mu_{u}=\mu_{d}=\mu_{s}=\mu_{B}/3$~\cite{Schaefer:2008hk,sbet-PLSM}.  Neglecting the small difference in masses of light quarks, the $m_{fl}$  for non-strange and strange quarks  can be given by~\cite{Kovacs:2006ym}
\begin{eqnarray}
m_q = g \frac{\sigma_x}{2}, \label{qmass}\quad m_s = g \frac{\sigma_y}{\sqrt{2}}.  \label{sqmass}
\end{eqnarray} 

In order to obtain the $T$ and $\mu_{B}$ dependence of order parameters, $\sigma_x$, $ \sigma_y$, $\phi$ and $\phi^*$,  we minimize the thermodynamic potential, Eq.~(\ref{potential}), with respect to these mean variables, i.e.
\begin{eqnarray}\label{cond1}
\left.\frac{\partial \Omega}{\partial \sigma_x}= \frac{\partial
	\Omega}{\partial \sigma_y}= \frac{\partial \Omega}{\partial
	\phi}= \frac{\partial \Omega}{\partial \phi^*}\right|_{min} =0,
\end{eqnarray}
where $\sigma_{x}=\bar{\sigma}_{x},~\sigma_{y}=\bar{\sigma}_y,~\phi=\bar{\phi},~\phi^*=\bar{\phi}^*$ labels global minimum.
	\begin{table}
	\begin{center}
		\begin{tabular}{|c|c|c|c|}
			\hline
			$c$ (MeV) &$\lambda_{1}$ &  $m^{2} $ ($\mathrm{MeV^{2}}$) & $\lambda_{2} $  \\
			\hline
			$4807.84$ & $13.49$ & $- (306.26)^2$ &  $46.48$   \\
			\hline
			$h_{x} $ ($\mathrm{MeV^{3}}$)& $h_{y} $ ($\mathrm{MeV^{3}}$)&$m_{\sigma}$(MeV) & $$   \\
			\hline
			$(120.73)^3$ & $(336.41)^3$ & $800$ & $$ \\
			\hline
		\end{tabular} 
		\caption{The parameters of PLSM  employed in the present calculation. 
			\label{par_tab1}}
	\end{center}
\end{table}
\subsection{Hadron masses}
 We firstly present the procedure of calculating $T$ and $\mu_{B}$ dependent masses of the pseudo-scalar $(\pi$, $\eta$, $\eta'$, $K$) and scalar ($\sigma$, $a_{0}$, $f_{0}$, $\kappa$) mesons in the framework of PLSM. In thermal field theory, the scalar and pseudo-scalar meson masses are defined by  the second derivative of the temperature and quark chemical potential dependent thermodynamic   potential $\Omega(T,\mu_{fl})$ with respect to corresponding scalar fields  
$\alpha_{S,x}=\sigma_{x}$ and pseudo-scalar fields $\alpha_{P,x}=\pi_{x} 
(x, y=0,...,8)$, which can be expressed as~\cite{Schaefer:2008hk}
\begin{equation}~\label{eq:mxy}
m_{i,xy}^{2}\big|_{T} = \frac{\partial^{2}\Omega(T,\mu_{fl})}
{\partial\alpha_{i,x}\partial\alpha_{i,y}}\big|_{min}=(m_{\alpha,xy}^{\mathrm{m}})^{2}+(\delta m_{\alpha,xy}^{T})^{2},
\end{equation}
where \textit{min} denotes to minimize the grand potential {\color{red}} and $i=S(P)$ corresponds to the scalar (pseudo-scalar) mesons.
 The first term  in Eq.~(\ref{eq:mxy}) is  vacuum meson mass calculated from the second derivative of purely mesonic potential. The second term corresponds to  in-medium modification of meson mass due to   quark-antiquark  potential at finite temperature and baryon chemical potential, which can be given as
\begin{eqnarray}\label{eq:deltamxy}
(\delta m_{i,xy}^{T})^{2}&=&3\sum_{fl=u,d,s}\int\frac{d^{3}p}{(2\pi)^{3}}\frac{1}{E_{fl}}[(A_{fl}^{+}+A_{fl}^{-})\nn\\
&&\times(m_{fl,xy}^{2}-\frac{m_{fl,x}^{2}m_{fl,y}^2}{2E_{fl}^2})\nn\\&&+(B_{fl}^{+}+B_{fl}^{-})(\frac{m_{fl,x}^{2}m_{fl,y}^2}{2E_{fl}T})].
\end{eqnarray}
The squared constituent quark mass derivative with respect to the meson field $\alpha_{i,a}$, $m_{fl,a}\equiv\partial m^2_{fl}/\partial\alpha_{i,a}$, and with respect to meson fields $\alpha_{i,a}\alpha_{i,b}$, $m_{fl,ab}\equiv\partial m^2_{fl}/\partial\alpha_{i,a}\partial\alpha_{i,b}$, can be taken from Table III in  Ref.~\cite{Schaefer:2008hk}.
The notations $A_{fl}^{\pm}$ and $B_{fl}^{\pm}$ in Eq.~(\ref{eq:deltamxy})   have the following definitions,
\begin{eqnarray}
A_{fl}^{+}=\frac{\phi e^{-E_{fl}^{+}/T}+2\phi^*e^{-2E_{fl}^{+}/T}+e^{-3E_{fl}^{+}/T}}{g_{fl}^{+}},\\
A_{fl}^{-}=\frac{\phi^* e^{-E_{fl}^{-}/T}+2\phi e^{-2E_{fl}^{-}/T}+e^{-3E_{fl}^{-}/T}}{g_{fl}^{-}},
\end{eqnarray}
and $B_{fl}^{\pm}=3(A_{fl}^{\pm})^2-C_{fl}^\pm$, where $C_{fl}^{\pm}$ is defined as
\begin{eqnarray}
C_{fl}^{+}=\frac{\phi e^{-E_{fl}^{+}/T}+4\phi^*e^{-2E_{fl}^{+}/T}+3e^{-3E_{fl}^{+}/T}}{g_{fl}^{+}},\\
C_{fl}^{-}=\frac{\phi^* e^{-E_{fl}^{-}/T}+4\phi e^{-2E_{fl}^{-}/T}+3e^{-3E_{fl}^{-}/T}}{g_{fl}^{-}}.
\end{eqnarray}
 
 Then the  squared masses of four scalar mesons are given as~\cite{Schaefer:2008hk} 
\begin{eqnarray}
m_{a_{0}}^2&=&(m_{a_{0}}^\mathrm{m})^2+(\delta m_{S,11}^T)^2,\label{eq:mao}
\\
m_{\kappa}^2&=&(m_{\kappa}^\mathrm{m})^2+(\delta m_{S,44}^T)^2,
\\
m_{\sigma}^2&=&m_{S,00}^2\cos^2\theta_{S}+m_{S,88}^2\sin^2\theta_{S}\nn\\
&&+2m_{S,08}^2\sin\theta_{S}\cos\theta_{S},\\
m_{f_{0}}^2&=&m_{S,00}^2\sin^2\theta_{S}+m_{S,88}^2\cos^2\theta_{S}\nn\\
&&-2m_{S,08}^2\sin\theta_{S}\cos\theta_{S}.
\end{eqnarray}
And the four pseudo-scalar meson masses are
\begin{eqnarray}
m_{\pi}^2 &=& (m_{\pi}^{\mathrm{m}})^2+(\delta m_{P,11}^T)^2,\\
m_{K}^2&=&(m_{K}^{\mathrm{m}})^2+(\delta m_{P,44}^T)^2,\\
m_{\eta^{'}}^2&=& m_{P,00}^2\cos^2\theta_{P}+m_{P,88}^2\sin^2\theta_{P}\nn\\&&+2
m_{P,08}^2\sin\theta_{P}\cos\theta_{P},\\
m_{\eta}^2&=& m_{P,00}^2\sin^2\theta_{P}+m_{P,88}^2\cos^2\theta_{P}\nn\\&&-2
m_{P,08}^2\sin\theta_{P}\cos\theta_{P},\label{eq:meta}
\end{eqnarray}
where the mixing angles $\theta_{S(P)}$ is given by 
\begin{eqnarray}
\tan2\theta_{i}=(\frac{2m_{i,08}^2}{m_{i,00}^2-m_{i,88}^2}),\quad i=S,P.
\end{eqnarray}
and $m_{i,00/88/08}^2=(m_{P,00/88/08}^\mathrm{m})^2+\delta(m_{P,00/88/08}^{T})^2$.
 The detailed expressions  of vacuum contributions   ($(m_{a_{0}}^\mathrm{m})^2$, $(m_{\kappa}^\mathrm{m})^2$, $(m_{\pi}^{\mathrm{m}})^2$, $(m_{K}^{\mathrm{m}})^2$ and $(m_{i,00/88/08}^\mathrm{m})^2$) from purely mesonic potential  in Eqs.~(\ref{eq:mao})-(\ref{eq:meta}) can be obtained from Refs.~\cite{Tawfik:2014gga,Schaefer:2008hk}. 
 \begin{figure}
 	\includegraphics[width=2.5in,height=3in]{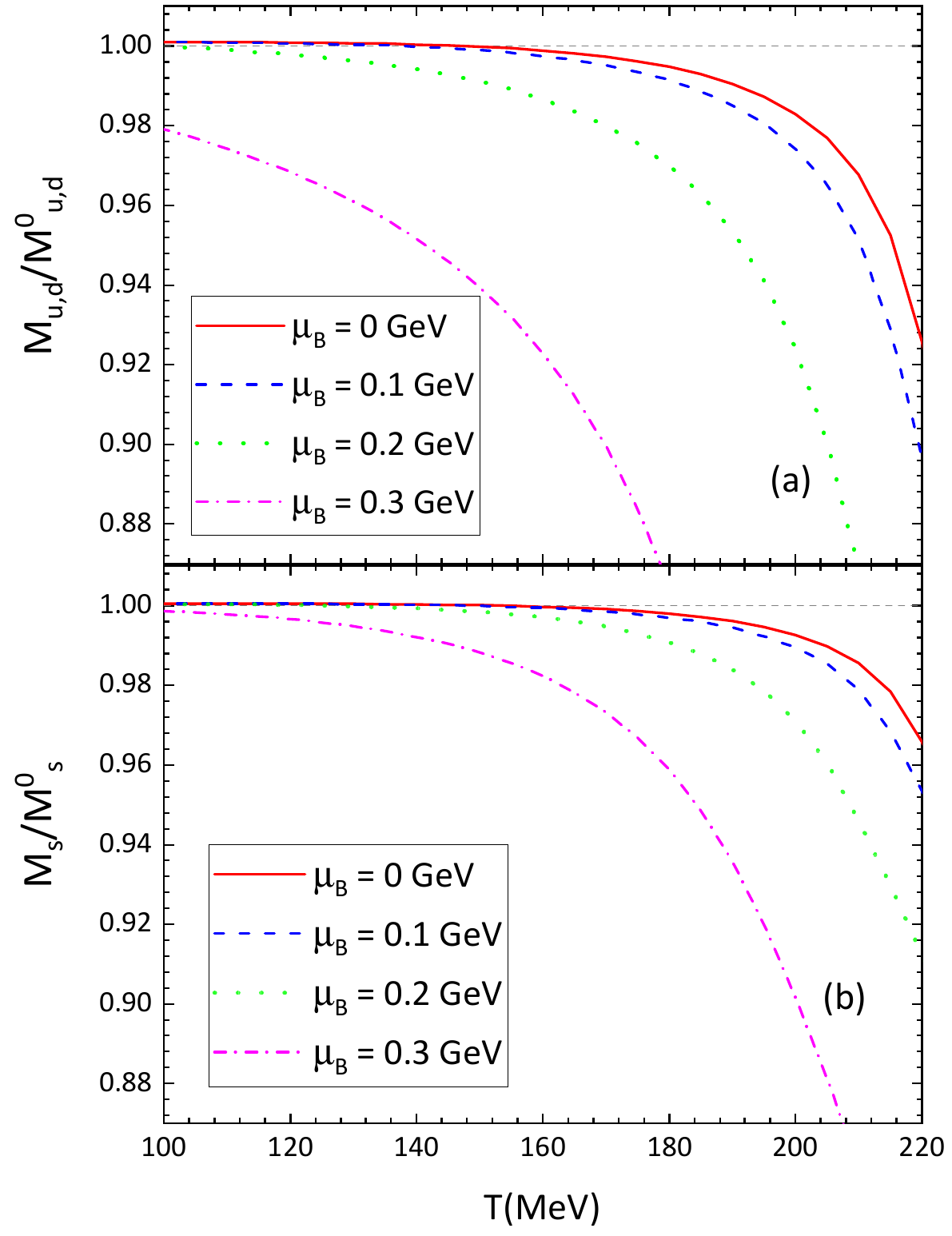}
 	\caption{\label{fig:epsart}(Color online) (a) The temperature dependence 
 		of the normalized light ($u,d$) constituent quark mass ($M_{u,d}/M^0_{u,d}$) for $\mu_{B}=0$ 
 		(red solid line), 0.1 GeV (blue dashed line), 0.2 GeV (green dotted line) 
 		and 0.3 GeV (purple dash-dotted line). 
 		(b) The temperature dependence of the normalized strange constituent quark 	mass ($M_{s}/M^0_{s}$) for different $\mu_{B}$. The light and strange vacuum 	constituent quark masses are taken as $M_{u,d}^{0}=300$ MeV and 
 		$M_{s}^{0}=433$ MeV~\cite{Schaefer:2008hk} in PLSM, respectively.}
 \end{figure}
 
Next, for all baryons and other heavier mesons, the dependence of their masses on  $T$ and $\mu_{B}$  can be obtained by introducing  a generalized  
scaling rule~\cite{Kadam:2015fza,Leupold,Jankowski:2012ms,Blaschke:2011hm,Blaschke:2015nma}, which
assumes that hadron masses are linear in the constituent quark masses, 
\begin{eqnarray}\label{eq:MBM}
M_{B/M}(T,\mu_{B})&=&M_{B/M}(0,0)+(N_{q}-N_{s})\delta 
M_{q}(T,\mu_{B})\nonumber\\
&&+N_{s}\delta M_{s}(T,\mu_{B}),
\end{eqnarray}
where the subscript $B/M$ stands for a given baryon/meson, $M_{q/s}$ is the light/strange constituent quark mass. $\delta M_{q/s}$ in Eq.~(\ref{eq:MBM}) denotes the variation of the constituent quark mass with temperature and baryon chemical potential. $N_{q}$ and $N_{s}$ are the number of total quarks and strangeness content  in a given hadron, respectively. For the open strange hadrons, $N_{s}$ is simply the number of strange (antistrange) quarks. For hidden strange  mesons, $N_{s}=2/3$ for the flavor singlet and $N_{s}=4/3$ for the flavor octet. However, we bear in mind that  this approach obtaining thermal hadron masses  is sketchy, and still needs  improvement in the future.
Fig.~\ref{fig:epsart} shows the normalized light constituent quark mass $M_{u,d}/M_{u,d}^{0}$ 
and the normalized   strange constituent quark mass $M_{s}/M_{s}^{0}$ as a function of  $T$ for different fixed $\mu_{B}$ in PLSM. The temperature behavior of the normalized constituent quark mass shows a smoothly decreasing feature.  The initial temperature at which the light/strange constituent quark masses begin to melt is  $T\sim 160/180$ MeV (not the chiral pseudo-critical temperature) for $\mu_{B}=0$ GeV. As  $\mu_{B}$ grows, $M_{u,d/s}/M_{u,d/s}^{0}$ begins to decrease at  smaller temperature  and the decreasing feature of constituent quark masses becomes more prominent.

\begin{figure*}
  	\includegraphics[width=6in,height=4.in]{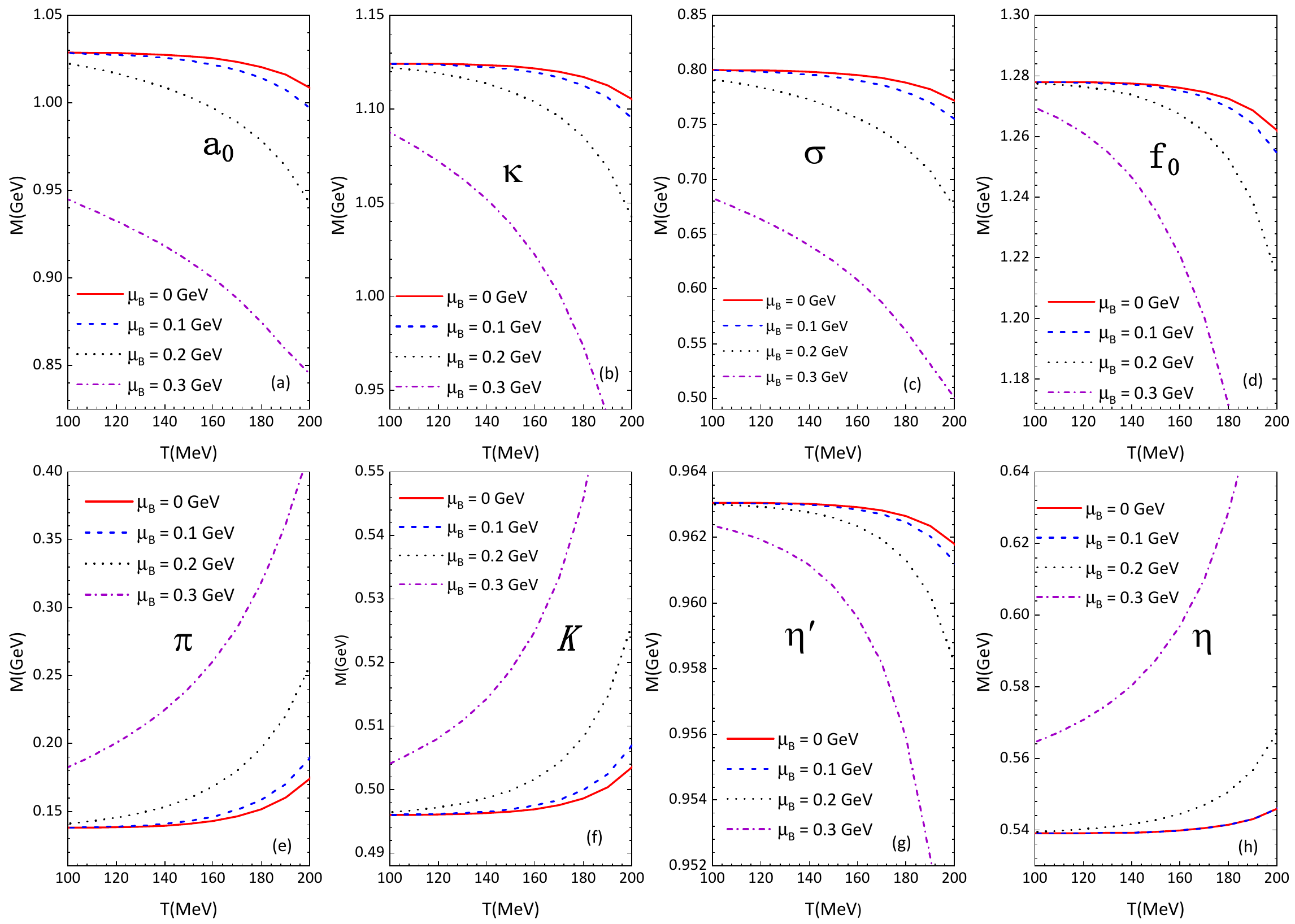}
  	\caption{\label{fig:meson} (Color online) The temperature dependencies of the  scalar mesons $a_{0}$,~$\kappa$,~$\sigma$,~$f_{0}$ and the pseudo-scalar 
  	mesons $\pi$,~$K$,~$\eta'$,~$\eta$ at $\mu_{B}=0$ (solid red lines), $0.1$~GeV (dashed blue lines), $0.2$~GeV (dotted black lines) and $0.3$~GeV (dotted-dashed purple lines)  in the framework of PLSM.}
\end{figure*}

Fig.~\ref{fig:meson} shows the temperature and baryon chemical potential dependencies of the pseudo-scalar mesons ($\pi$, $K$, $\eta'$, $\eta$) and scalar mesons ($a_{0}, \kappa, \sigma, 
f_{0})$ in  PLSM. The masses of these states degenerate at $T \sim 160$ MeV for $\mu_{B}=0$ and $0.1$ GeV cases. For $\mu_{B}=0.2$ and $0.3$ GeV cases, these states degenerate at $T \sim 130$ MeV and $T<100$ MeV, respectively. Therefore the melting behavior of hadron masses can quantitatively  affect the thermodynamic quantities and transport coefficients of hadronic matter, which will be seen later in  Sec.~\ref{sec:results}.


\section{Transport coefficients}
\label{sec:transport-coefficients}

Transport coefficients in the medium composed of quasi-particles whose  masses depend on temperature and chemical potential can be derived by utilizing the relativistic kinetic theory under relaxation time approximation~\cite{Chakraborty:2010fr,Mitra:2018akk,sbe-PQM model}. The  general expressions of shear viscosity ($\eta$), electrical conductivity ($\sigma_{el}$) and  thermal conductivity  ($\lambda$)  can be written as~\cite{Chakraborty:2010fr,sbe-PQM model}
\begin{eqnarray}
\eta&=&\frac{1}{15T}\sum_{i}^{}g_{i}\int\frac{d^{3}p}{(2\pi)^{3}} 
\frac{p^{4}}{E_{i}^{2}}\tau_{i}f_{i}^{id}(1\pm f_{i}^{id}),\label{eq:eta}\\
\sigma_{el}&=&\frac{1}{3T}\frac{4\pi}{137}\sum_{i}^{}g_{i}e_{i}^{2}\int\frac{d^{3}p}
{(2\pi)^{3}}\frac{p^{2}}{E_{i}}\tau_{i}f_{i}^{id}(1\pm f_{i}^{id}),\label{eq:sigma}\\
\lambda&=&\left(\frac{w}{n_{B}T}\right)^{2}\sum_{i}^{}g_{i}\int
\frac{d^{3}p}{(2\pi)^{3}}\frac{p^{2}}{3E_{i}^{2}}\tau_{i}
\left(B_{i}-\frac{n_{B}E_{i}}{w}\right)^{2}\label{eq:heat} \nonumber \\  
&& \times f_{i}^{id}(1\pm f_{i}^{id}).
\end{eqnarray}
Here $e_{i}$ and $B_{i}$ are the electric charge and baryon number of hadron species $i$, respectively. $w$ is total enthalpy density. The sign $\pm$ corresponds to bosons and fermions respectively.
$\tau_{i}$ is the thermal  relaxation time of hadron species $i$. We assume only elastic scattering between hadrons, so inverse relaxation time $\tau_{i}^{-1}$  for the collision process of $i(p_{1})+j(p_{2})\rightarrow i(p_{3})+j(p_{4})$ 
can be given by~\cite{tau}
\begin{eqnarray}\label{eq:tau}
  \tau_{i}^{-1}=\sum_{j}^{}\frac{g_{j}}{1+\delta_{ij}}\int\prod_{k=2}^{4}\frac{d\Gamma_{k}}{2E_{1}}
  (2\pi)^{4}\delta^{4}(P_{tot})|\bar{M}|^{2}f_{j}(p_2)\nonumber,\\
\end{eqnarray}
where $d\Gamma_{k}=d^{3}p_{k}/(2\pi)^{3}/(2E_{k})$,  
$\delta^4(P_{tot})=\delta^3(p_{1}+p_{2}-p_{3}-p_{4})\delta(E_1+E_{2}-E_{3}-E_{4})$, the factor $1/(1+\delta_{ij})$ is to avoid double counting for idential incoming particle species. In Eq.~(\ref{eq:tau}), the average of the 
initial degeneracy factor and the sum of final degeneracy factor are   implicitly included in the matrix element $(\bar{M})$. Using the formula of scattering cross section~\cite{peskin}
\begin{eqnarray}
  \sigma_{ij}=\frac{\int\prod_{k=3}^{4}  
  d\Gamma_{k}(2\pi)^{4}\delta^{4}(P_{tot})|\bar{M}|^{2}}{4\sqrt{(P_{1}\cdot
  		P_{2})^{2}-m_{i}^{2}m_{j}^{2}}},
\end{eqnarray}
with four-momentum $P_{1(2)}=(E_{1(2)},p_{1(2)})$
then we can rewrite $\tau_{i}^{-1}$ and take thermal averaging
\begin{eqnarray}
  \tau_{i}^{-1}\equiv\sum_{j}^{}\frac{n_{j}}{1+\delta_{ij}}\langle\sigma_{ij} 
  v_{ij}\rangle,
\end{eqnarray}
where $n_{j}=g_{j}\int d^{3}p_{2}/(2\pi)^{3}f_{j}(p_2)$ is the number density of particle species $j$. It is important to note that if partice species $j$ is a baryon/antibaryon, 
the detailed form of the number density can be modified in 
van der Waals hadron resonance gas,
\begin{equation}
\label{equ:kappa-r-region}
n_{j}(T,\mu_{j})=\left\{\begin{array}{l}
n_{j}^{id}(T,\mu_{j}) \ , \quad \text{in ideal HRG} \ ;\\
 F(h_{B(\bar{B})})n_{j}^{id}(T,\mu_{j}^{B(\bar{B})*})\ ,~\text{in VDW HRG} \ 
 .
\end{array}\right.
\end{equation}
 The Lorentz scalar flow factor is defined as
\begin{eqnarray}
  v_{ij}=\frac{\sqrt{(P_{1}\cdot 
  		P_{2})^{2}-m_{i}^{2}m_{j}^{2}}}{E_{1}E_{2}}.
\end{eqnarray}
Therefore the thermal average cross section with Maxwell-Boltzmann distribution approximation after some uncomplicated simplification can be written as the following form  
\begin{eqnarray}
\langle\sigma_{ab} 
  v_{ij}\rangle&=&\frac{\int 
  	d^{3}p_{1}d^{3}p_{2}f_{i}^{id}(p_1)f_{j}^{id}(p_2)\sigma_{ij}v_{ij}}{\int
  	d^{3}p_{1}d^{3}p_{2}f_{i}^{id}(p_1)f_{j}^{id}(p_2)}\nonumber\\
  &=&\frac{\int 
  	d^{3}p_{1}d^{3}p_{2}e^{-E_{1}\beta}e^{-E_{2}\beta}\sigma_{ij}v_{ij}}{\int 	d^{3}p_{1}d^{3}p_{2}e^{-E_{1}\beta}e^{-E_{2}\beta}}\nonumber\\
  &=&\frac{\beta\int_{S_{0}}^{\infty}\sigma_{ij}\gamma(S)K_{1}(\beta\sqrt{S})\frac{1}{2\sqrt{S}}dS}{4m_{i}^{2}m_{j}^{2}K_{2}
  	(\beta m_{i})K_{2}(\beta m_{j})},
\end{eqnarray}
where $\sqrt{S}$ is center-of-mass energy, $S_{0}=(m_{i}+m_{j})^{2}$,   
$\gamma(S)=[S-(m_{i}+m_{j})^{2}][S-(m_{i}-m_{j})^{2}]$.   
$K_{n}$ is the modified Bessel function of order $n$. In this work we regard all hadrons  as hard spheres which have the same radius $r_{h}$ as nucleons, so $\sigma_{ij}$ is a constant with $\sigma_{ij}=4\pi r^{2}_{h}$.

\section{Numerical Results and Discussions}
\label{sec:results}
In the following, we consider an extension of VDWHRG model by including thermal evolution of hadron masses, and refer to this new model as Thermal VDWHRG (TVDWHRG) model. In the treatment of HRG model   we include all hadrons and resonances up to 2.0 GeV listed in the Particle Data Group Book of 2014~\cite{PDG2014}.  
The VDW model yields $a\approx 239~{\rm MeV~fm^{3}}$ and $b=\frac{4\pi r^{3}_{n}}{3}\approx 3.42~{\rm fm}^{3}$ ($r_{n}$ is the radius of 
nucleons) from fitting  the properties of nuclear matter at zero   
temperature~\cite{VVgroundstate}.
 
\begin{figure*}
   	\includegraphics[width=4.2in,height=4.5in]{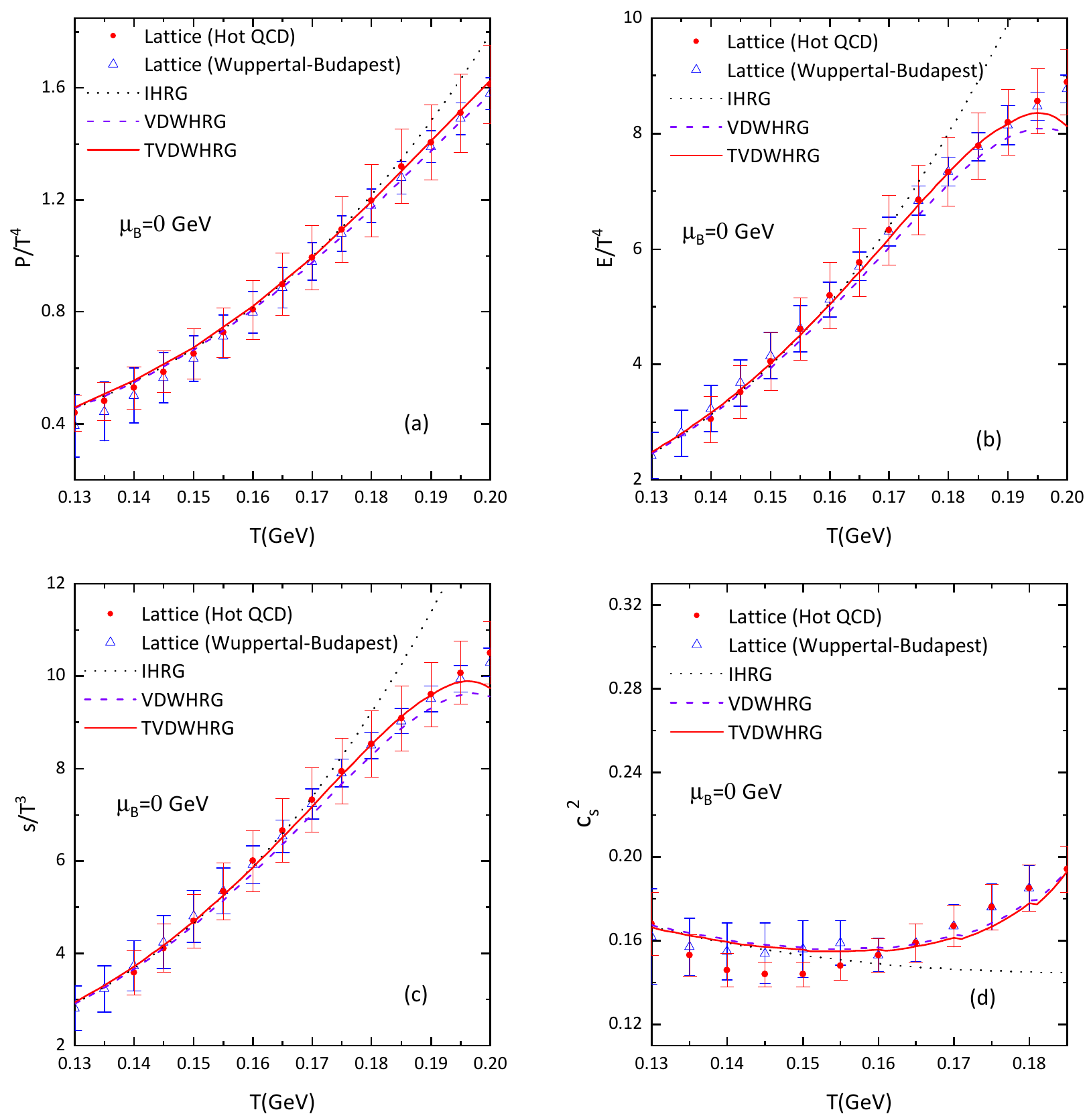}
   	\caption{\label{fig:PSECS} (Color online) The temperature dependencies of 
   	different thermodynamics within IHRG model (black dashed lines), VDWHRG 
   	model(blue dashed lines) and TVDHRG model (red solid lines) at $\mu_{B}=0$. 
   	The Lattice QCD results are taken from the 
   	Wuppertal-Budapest~\cite{Borsanyi:2013bia} (red solid circle symbol with 
   	error bar) and the HotQCD collaborations~\cite{lattice-u} (blue uptriangle 
   	symbol with error bar).}
 \end{figure*}

The temperature dependencies of the scaled pressure $P/T^{4}$, the scaled energy density $\epsilon/T^{4}$, the scaled entropy density 
$s/T^{3}$ and speed of sound squared $c_{s}^{2}=dP/d\epsilon$ at $\mu_{B}=0$ GeV within IHRG, VDWHRG and TVDWHRG models are depicted in Fig.~\ref{fig:PSECS}.
It is noted that in Fig.~\ref{fig:PSECS}($a$-$c$), comparing with the results in IHRG model, the pressure, the energy density and the entropy density within VDWHRG and TVDWHRG models have a modest suppression at $T>0.16$ GeV due to the suppression of the number density of (anti)baryons in the medium.
The scaled pressure $P/T^{4}$, the scaled energy density $\epsilon/T^{4}$ and the scaled entropy density $s/T^{3}$ in TVDWHRG model have a better agreement with the Lattice QCD data of the Wuppertal-Budapest~\cite{Borsanyi:2013bia} and the Hot QCD collaborations~\cite{lattice-u} up to $T=0.195$ GeV. 
The mild quantitative difference between these thermodynamics  in VDWHRG model and the counterparts in TVDWHRG  model at $T>0.16$ GeV results from an enhancement factor of $\exp[-m(T,\mu_{B})/T]$ with the decrease of hadron masses in TVDWHRG model. In Fig.~\ref{fig:PSECS} ($d$)  we observe that the speed of sound squared $c_{s}^{2}$ in TVWDHRG model or VWDHRG model is consistent with the Lattice QCD data at $T = 0.165\sim 0.18$ GeV. While $c_{s}^{2}$ within all considered HRG models gives a bad fit to the Lattice QCD data of the Hot QCD collaborations at $T=0.135 \sim 0.155$ GeV. 
In addition,  the pressure, the energy  density, the entropy density and speed of sound square are not sensitive to the choice of considered HRG models at $T < 0.16$ GeV. It can be explained in two aspects. (i) The VDW interactions between baryon-baryon  pairs and between  antibaryon-antibaryon pairs for $\mu_{B}=0$ GeV are relatively weak at  $T < 0.16$ GeV because at low $T$ the contribution of mesons is dominant in system compared to  the contribution of  (anti)baryons. (ii) At $T < 0.16$ GeV, the masses of hadrons for $\mu_{B}=0$ are nearly not affected by temperature, as seen from  Fig.~\ref{fig:epsart} and Fig.~\ref{fig:meson}.

   \begin{figure*}
 	\includegraphics[width=6.2in,height=2.3in]{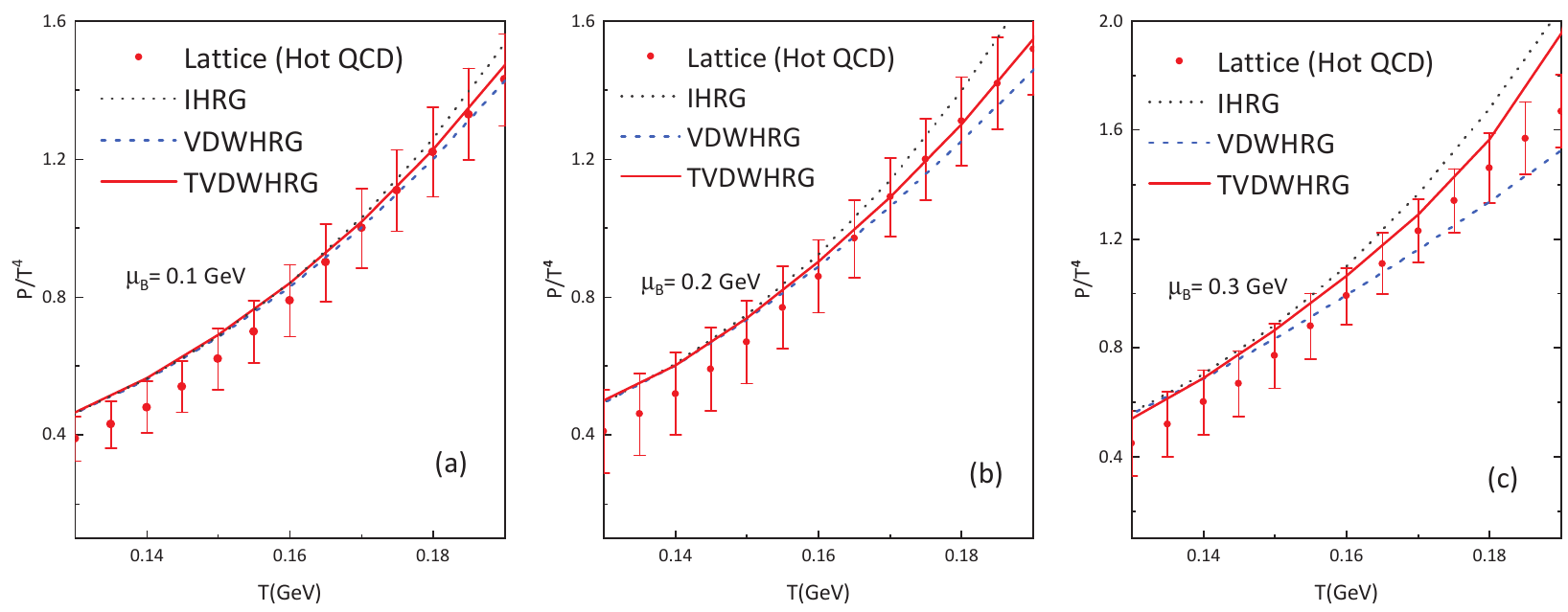}
 	\caption{\label{fig:P} (Color online) The temperature dependence of the 
 	scaled pressure ($P/T^{4}$) within IHRG model (dashed black lines), VDWHRG 
 	model(wide dashed purple lines) and TVDWHRG model (solid red lines) at 
 	$\mu_{B}=0.1 (a)$, $0.2 (b)$ and $0.3$ GeV $(c)$. The Lattice QCD results 
 	(red symbol with error bar) are taken from Ref.~\cite{latticeu2}.}
 \end{figure*}

Fig.~\ref{fig:P}($a$-$c$) shows the scaled pressure  as a function of temperature for $\mu_{B}=0.1$, $0.2$ and $0.3$ GeV. 
We see that the scaled pressure is underestimated by all models at $T< 0.16$ GeV.
For the cases of  $\mu_{B}=0.1$ and $0.2$ GeV, the scaled pressure within TVDWHRG model fits better with the Lattice QCD data at $T= 0.16\sim0.19$ GeV than that within VDWHRG model or IHRG model. Compared to VDWHRG model, the scaled pressure for zero and small  $\mu_{B}$ ($viz$, $\mu_{B}=0.1$ and $0.2$ GeV) in TVDWHRG model have a small quantitative enhancement.
At  higher $\mu_{B}$ (i.e. $\mu_{B}=0.3$~GeV) we notice that the scaled pressure within TVDWHRG model  is significantly higher than that within VDWHRG model, as shown in Fig.~\ref{fig:P}($c$). This means that with the  increase of $\mu_{B}$, the effect of thermal hadron masses on thermodynamics becomes more influential. However, the scaled pressure for $\mu_{B}=0.3$ GeV fails to simulate the  Lattice QCD data within all considered HRG models. 
There are two possible reasons for the failure: (i) The parameters of VDW model may vary with $\mu_{B}$~\cite{Sarkar:2018mbk}. (ii) It is a challenging task for the  Lattice QCD simulation to give very reliable predictions of these quantities due to so-called sign problem at nonzero $\mu_{B}$. The Lattice QCD data we used here is only estimated up to $\mu_{B}^{2}$~\cite{latticeu2}.
 Therefore, in the case of nonzero baryon chemical potential, we may not pay much attention to comparing our results with the Lattice QCD data in precision, instead we explore the effects of thermal hadron masses and VDW interactions on  thermodynamic  quantities and transport coefficients in hot hadronic matter.
  
\begin{figure}
  	\includegraphics[width=2.4in,height=2.5in]{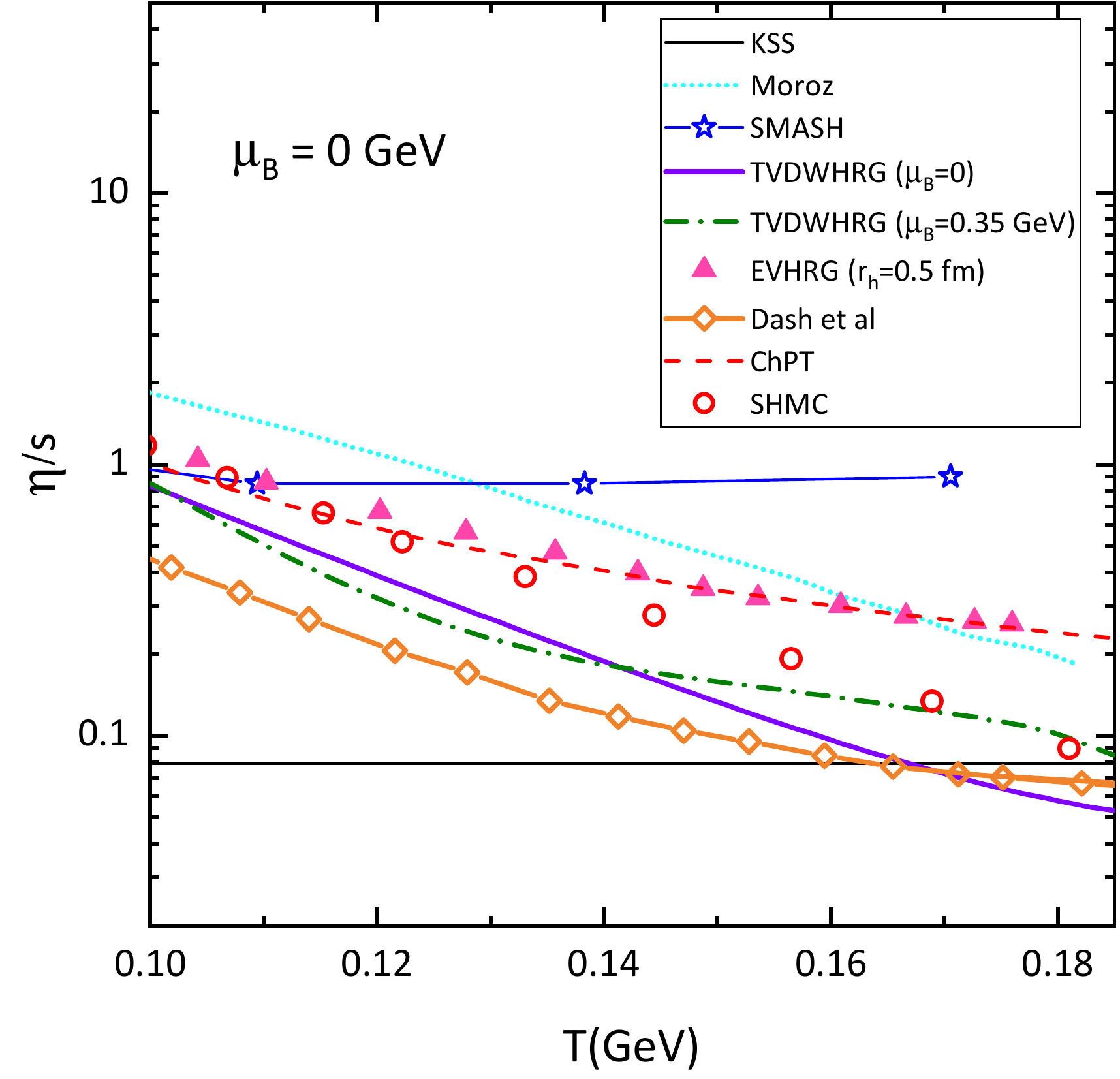}
  	\caption{\label{fig:Eta/s-vs}(Color online) Comparison of several calculations for the shear viscosity to entropy density ratio $\eta/s$ at $\mu_{B}=0$~GeV, see text for more explanations.}
  \end{figure}

The temperature dependence of shear viscosity to entropy density ratio, $\eta/s$,  within TVDWHRG model at $\mu_{B}=0$~GeV (purple solid line) is shown in Fig.~\ref{fig:Eta/s-vs}. We can note that  $\eta/s$ decreases with increasing temperature  and the value of $\eta/s$ within TVDWHRG model  meets the quantum lower bound, $\eta/s=1/4\pi$,  proposed by  Kovtun, Son and Starinets (KSS)~\cite{kss}  in the vicinity of $T=0.17$~GeV.
   This means the applicability of TVDWHRG model  should be restricted to the temperature  domain in which  $\eta/s>1/4\pi$. 
    At $\mu_{B}=0.35$~GeV, our result of $\eta/s$ (green dotted-dashed line) remains above the KSS bound in entire temperature domain we considered here.  We also notice that in low $T$ domain, $\eta/s$   is slightly smaller at $\mu_{B}=0.35$~GeV than at $\mu_{B}=0$~GeV, however, in high  $T$ domain  $\eta/s$  is higher  at $\mu_{B}=0.35$~GeV than   at $\mu_{B}=0$~GeV. This behavior is also observed in Fig.7 of Ref.~\cite{shear-CE2} and in Fig.6 of Ref.~\cite{Ghosh:2015lba}, we will discuss this behavior later.

Fig.~\ref{fig:Eta/s-vs} also demonstrates the comparison of our calculations with the  results from other related models for $\mu_{B}=0$~GeV. The open red circles  correspond to the result of $\eta/s$ for hadron phase using RTA within SHMC model~\cite{SHMC}.
  The  blue star-line corresponds to the result of $\eta/s$ for hadron gas using Kubo-Green formalism  in  SMASH transport code~\cite{shear-SMASH}.
   The result of $\eta/s$ by Dash {\it et al} (orange diamond-dashed line) is computed in the framework of an $S$-matrix based HRG model using  CE approximation and $K$-Matrix cross sections ~\cite{shear-CE2}. The red dotted line represents the result of $\eta/s$ using Green-Kubo formalism in unitarized ChPT which is a low-energy effective model of QCD describing the dynamics of the Nambu-Goldstone bosons~\cite{sek-cpt-pion1}. 
   The result of  $\eta/s$ by  Moroz (artic short dotted line) is  obtained  from solving  Boltzmann equation in RTA  while the cross sections are extracted from UrQMD~\cite{RTA3}.  The pink triangles show the calculation of  $\eta/s$ for hadron gas in EVHRG model~\cite{shear-evhrg4}. 
 All aforementioned works except  SMASH model give qualitative results which are similar with ours, although the exact magnitude of $\eta/s$ differs in different model estimations. 
   The result of Moroz~\cite{RTA3} is about 3 times larger than ours mainly due to the discrepancies in cross sections. In  Moroz's calculation, the cross sections extracted from UrQMD model  for different  hadron-hadron elastic collisions are different whereas  an overall constant cross section is used in current work.  The SHMC result~\cite{SHMC} is close to ours at $T<0.12$~GeV and is about 2 times larger than our estimation at $T>0.12$~GeV. This is mainly because although the hadron masses in SHMC model and TVDWHRG model are in-medium dependent, 
   the cross sections in SHMC model are temperature dependent rather than a constant.
   The result of Dash {\it et al}~\cite{shear-CE2}   is a factor of 2 smaller than ours at $T<0.14$~GeV, however, as  temperature increases further their result is very close to ours. The quantitative difference can be attributed  to the uses of various  approximation methods and cross sections. In our work the transport coefficients are calculated in RTA which is different to CE method. 
    We emphasize that in current work  the cross sections  are taken as constant, which assumption could be improved in future's studies.
  Furthermore, $\eta/s$ calculated by Dash \textit{et al} also violates the KSS bound near the critical temperature taken from Ref.~\cite{Tc}. The estimation of  $\eta/s$ in SMASH~\cite{shear-SMASH} is close to ours at $T<0.12 $~GeV while as temperature increases the SMASH result remains almost constant. This behavior can be explained as follows: Firstly, in our work we only contain  elastic binary collisons between hadrons with  constant cross sections while the energy dependent cross sections and  hadron interactions dominated by resonance formation are included in SMASH. Secondly,  the effect of resonance lifetimes on the relaxation time is considered in SMASH, whereas we use the  thermal averaged relaxation time  which contains no feedback from the resonance lifetimes (zero decay width used in our work for resonances). 
    The result in EVHRG model~\cite{shear-evhrg4} and  ChPT result~\cite{sek-cpt-pion1}  match well with ours at  $T<0.12$~GeV, however, their estimations are about 3 times larger than ours at high $T$. The numerical difference between  ChPT result  and ours  might be due to the fact that  at high $T$ more  meson-baryon scatterings are included in TVDWHRG model while  only  $\pi$-$\pi$ scattering  is considered in  ChPT.  The deviation between  the estimation of $\eta/s$ in  EVHRG model and ours at high $T$ can mainly arise from that, in Ref.~\cite{shear-evhrg4} authors consider the repulsive interaction is related to  all hadrons with same radius ($r_{h}=0.5$~fm), whereas, in TVDWHRG model only VDW interactions between pairs of (anti)baryons are included and all hadrons  have same radius as nucleons. 
    Actually, the thermal mass effect is not obvious on $\eta/s$  within TVDWHRG model at zero $\mu_{B}$ or small $\mu_{B}$, while as $\mu_{B}$ increases this effect becomes more pronounced, which can be shown later in Fig.~\ref{fig:Eta/s}. 
      
     \begin{figure*}
  	\includegraphics[width=5.5in,height=2.5in]{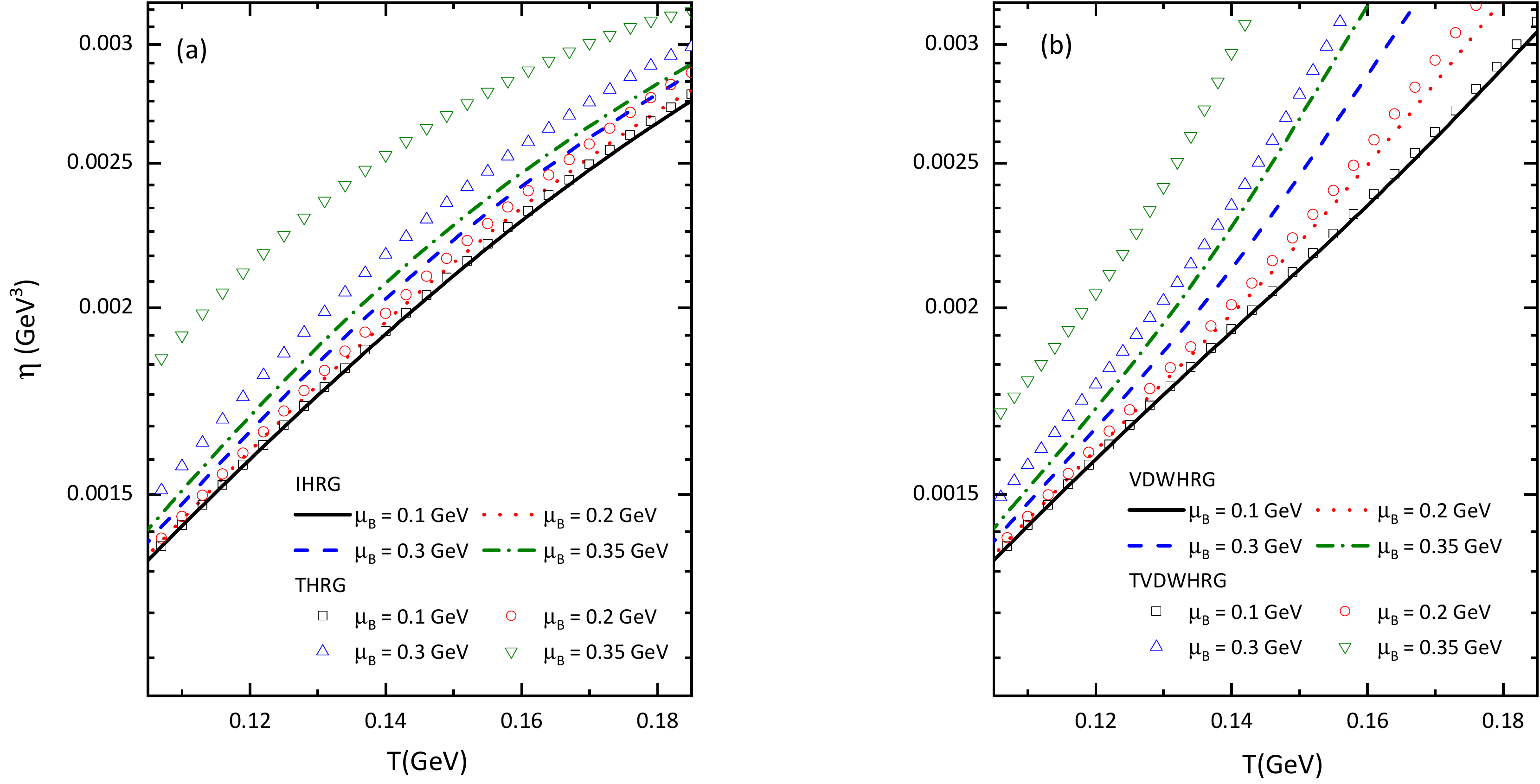}
  	\caption{\label{fig:Eta}(Color online) Left panel (a) shows the temperature dependence of shear viscosity $\eta$ within IHRG (lines) and THRG (symbols) model for $\mu_{B}=0.1$ (black solid line and square symbol), $0.2$ (red  dotted line and circular symbol), $0.3$ (blue dashed line and uptriangle symbol) and $0.35$ (green dashed-dotted line and downtriangle symbol) GeV. Right panel (b) shows the temperature dependence 
  	of $\eta$ within VDWHRG (lines) and TVDWHRG (symbols) models for $\mu_{B}=0.1$, $0.2$, $0.3$ and $0.35$ GeV.}
  \end{figure*}

Here we refer to hadron resonance gas model only including the effect of  thermal hadron masses as Thermal HRG (THRG) model. To better understand how  the effects of in-medium hadron masses and VDW interactions between  (anti)baryons  influence  transport coefficients in hadronic matter,
we compute the variation of transport coefficients with $T$ and $\mu_{B}$ in four HRG models: IHRG, THRG, VDWHRG, and TVDWHRG models.  The temperature dependence of shear viscosity $\eta$ at $\mu_{B}=0.1$,~$0.2$,~$0.3$,~$0.35$~GeV for all considered HRG models is depicted in Fig.~\ref{fig:Eta}. We observe that  $\eta$ within IHRG model increases monotonically as $T$ increases  at a fixed $\mu_{B}$.  This is because  the variation of shear viscosity $\eta$ with $T$ and $\mu_{B}$ in IHRG model mainly  comes from the number density in Eq.~(\ref{eq:eta}) rather than relaxation time.
  Alternatively, at a given $T$ the value of $\eta$ in IHRG model increases  as $\mu_{B}$ grows. 
 Considering the effect of thermal hadron  masses, the value of $\eta$ for $\mu_{B}=0.1$~GeV within THRG model  has a mild enhancement in relatively high $T$ domain as shown in Fig.~\ref{fig:Eta}($a$), which is similar to the result in Fig.3($a$) of Ref.~\cite{Kadam:2015fza}.  This is due to the fact that the number density has an enhancement by considering the effect of themal hadron masses.  As $\mu_{B}$ increases, the improvement of $\eta$ in THRG model is more obvious than in IHRG model, which arises from that the positive effect of thermal hadron masses on the number density strengthens significantly with increasing  $\mu_{B}$. 
 When the VDW interactions are taken into account in the estimation of $\eta$ (as in Fig.~\ref{fig:Eta}($b$)),   $\eta$ rises with  larger rate of increment at high temperature in VDWHRG model as compared to IHRG model.
 This  can be interpreted as follows: Firstly, at low $T$ the dominant contributions to total $\eta$ are light mesons which are nearly not affected by VDW interactions. Secondly, with increasing $T$ more and more baryons emerge in system and  the baryon density can be suppressed in VDWHRG model as compared to IHRG model.  However the relaxation times for all hadrons in  VDWHRG model have a significant enhancement  due to the scattering with baryons. As temperature increases, the effect of a rapid rise in the relaxation time wins over the impact of a fall in the number density within VDWHRG model. Furthermore, we see that  the evolution of  $\eta$ with $\mu_{B}$ in VDWHRG model mimics that in IHRG model.
   At high $\mu_{B}$ ($viz$, $\mu_{B}=0.3$ or $0.35$~GeV), considering  simultaneously the effects of VDW interactions and in-medium hadron masses, the number density in TVDWHRG model increases more sharply than that in VDWHRG model, although the relaxation time in TVDWHRG model is slightly reduced than that in VDWHRG model. The final result of the interplay of the number density and the relaxation time in  Eq.(\ref{eq:eta}) shows the TVDWHRG model give a further improvement in $\eta$ compared to VDWHRG model, as shown in Fig.~\ref{fig:Eta}($b$). 
   
    \begin{figure*}
	\includegraphics[width=5.5in,height=2.5in]{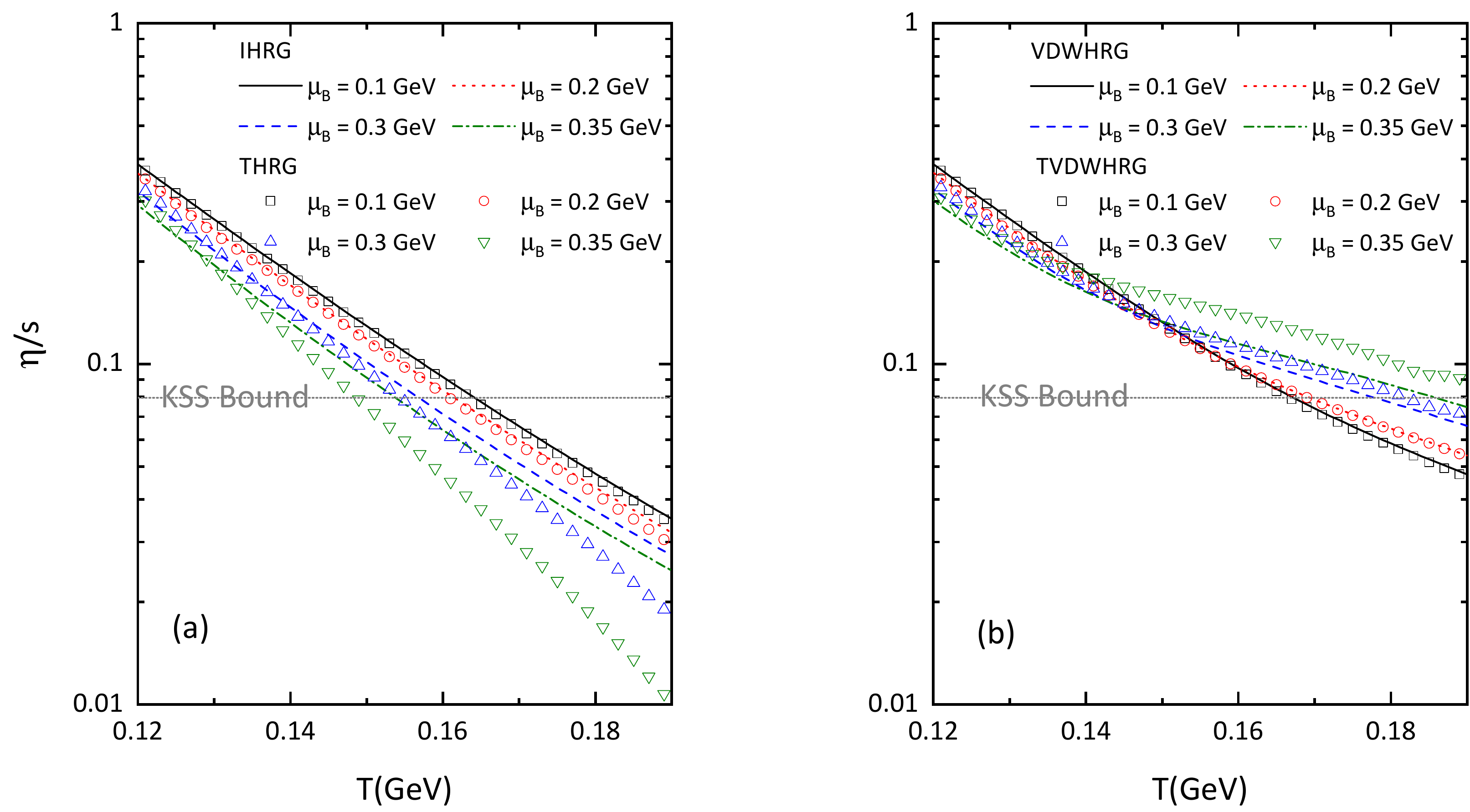}
	\caption{\label{fig:Eta/s}(Color online)  Same as Fig.~\ref{fig:Eta} for ratio $\eta/s$. The gray dotted line is KSS bound.}
\end{figure*}

Fig.~\ref{fig:Eta/s} ($a$) presents our calculation of  $\eta/s$ for  various $\mu_{B}$ in IHRG and THRG models. We note that the $T$ dependence of  $\eta/s$ within IHRG and THRG models is mainly governed by the inverse entropy density, $1/s$. The ratio $\eta/s$ in IHRG and THRG  models decreases as $T$ and $\mu_{B}$ increase solely due to the larger value of entropy density for high $T$ and high $\mu_{B}$.  Compared to IHRG model,  THRG model leads to a suppression of $\eta/s$, which arises from  the significant enhancement of the entropy density in THRG model. At small  $\mu_{B}$ ($\mu_{B}=0.1$ and 0.2~GeV) or zero $\mu_{B}$, $\eta/s$ is nearly unaffected by the inclusion of in-medium hadron masses. The reasons are twofold. On the one hand, the effect of thermal hadron masses   is weaker at small $\mu_{B}$ case than at high $\mu_{B}$ case. On the other hand, with the consideration of thermal hadron masses the increase in $\eta$  is nearly  neutralized by the decrease in $1/s$.
Hence,  from the quantitative aspect the effect of thermal hadron masses is important on $\eta/s$ especially at high $\mu_{B}$.

Fig.~\ref{fig:Eta/s}($b$) displays  $\eta/s$ in VDWHRG and TVDWHRG  models as a function of temperature at various $\mu_{B}$. 
 It is interesting to note that in VDWHRG or TVDWHRG model as $\mu_{B}$ grows $\eta/s$  decreases  at low $T$ whereas increases  at high $T$, which is qualitatively akin to the result of $\eta/s$  in Ref.~\cite{Ghosh:2015lba}.  This non-trivial behavior of $\eta/s$  in VDWHRG model case is not observed in EVHRG model~\cite{shear-evhrg4}. 
 The non-monotonous variation of  $\eta/s$ with $\mu_{B}$ is due to that in  high $T$ domain  as $\mu_{B}$ grows the rapid increase of $\eta$  (as in Fig.~\ref{fig:Eta}) greatly overwhelms the decrease of  $1/s$  in VDWHRG model.
Furthermore, at high $\mu_{B}$ ($viz$, $\mu_{B}=0.3$ and 0.35~GeV), the effect of VDW interactions on $\eta/s$ in high $T$ domain can be  strengthened further by the inclusion of thermal hadron masses, even though  thermal hadron masses itself have a negative effect on $\eta/s$.
Hence, the consideration of VDW interactions (thermal hadron masses) mainly changes  qualitatively (quantitatively)  the behavior of $\eta/s$.
In Fig.~\ref{fig:Eta/s}($b$) we also observe   the  location where  $\eta/s(T,\mu_{B})\simeq1/4\pi$  shifts toward  higher temperature with increasing  $\mu_{B}$ in TVDWHRG and VDWHRG models, contrary to  IHRG and THRG models case.

  \begin{figure}
	\includegraphics[width=2.4in,height=2.5in]{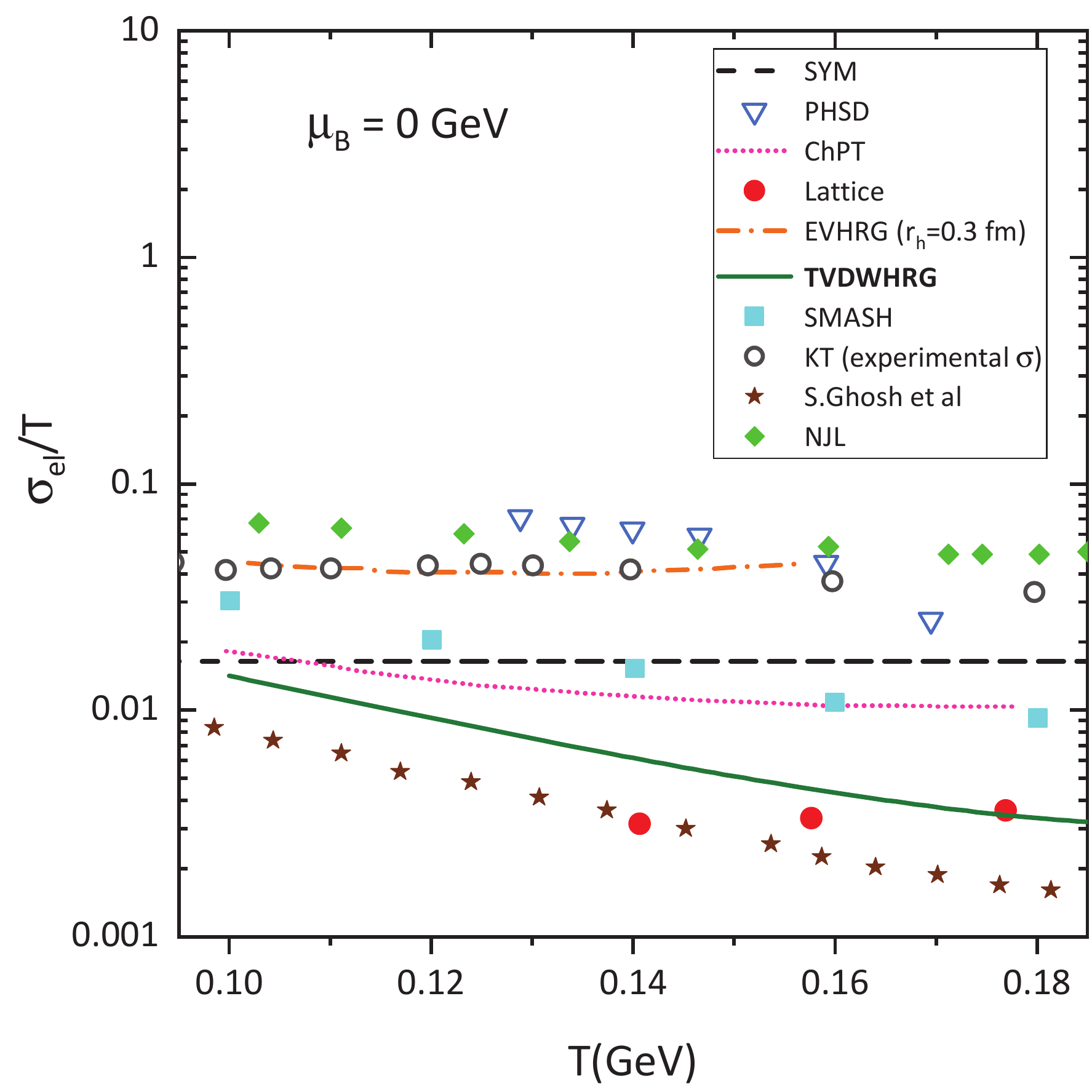}
	\caption{\label{fig:sigma-vs}(Color online) Comparison of several calculations for the scaled electrical conductivity $\sigma_{el}/T$ at $\mu_{B}=0$~GeV, see text for more explanations.}
\end{figure}

   \begin{figure*}
	\includegraphics[width=5.5in,height=2.5in]{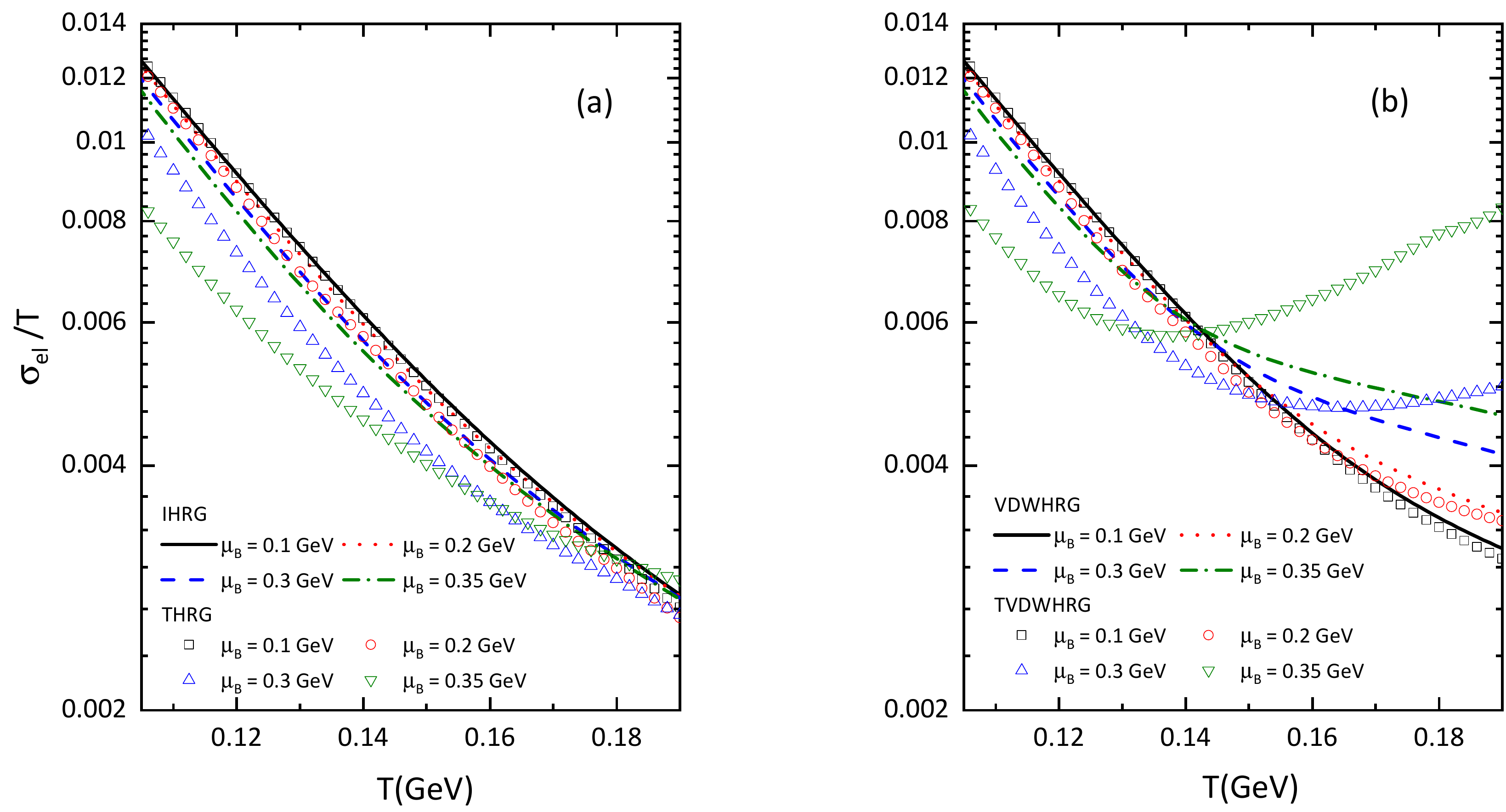}
	\caption{\label{fig:sigma}(Color online)  Same as Fig.~\ref{fig:Eta} for the 	scaled electrical  conductivity $\sigma_{el}/T$.}
\end{figure*}

  \begin{figure}
  	\includegraphics[width=2.5in,height=2.5in]{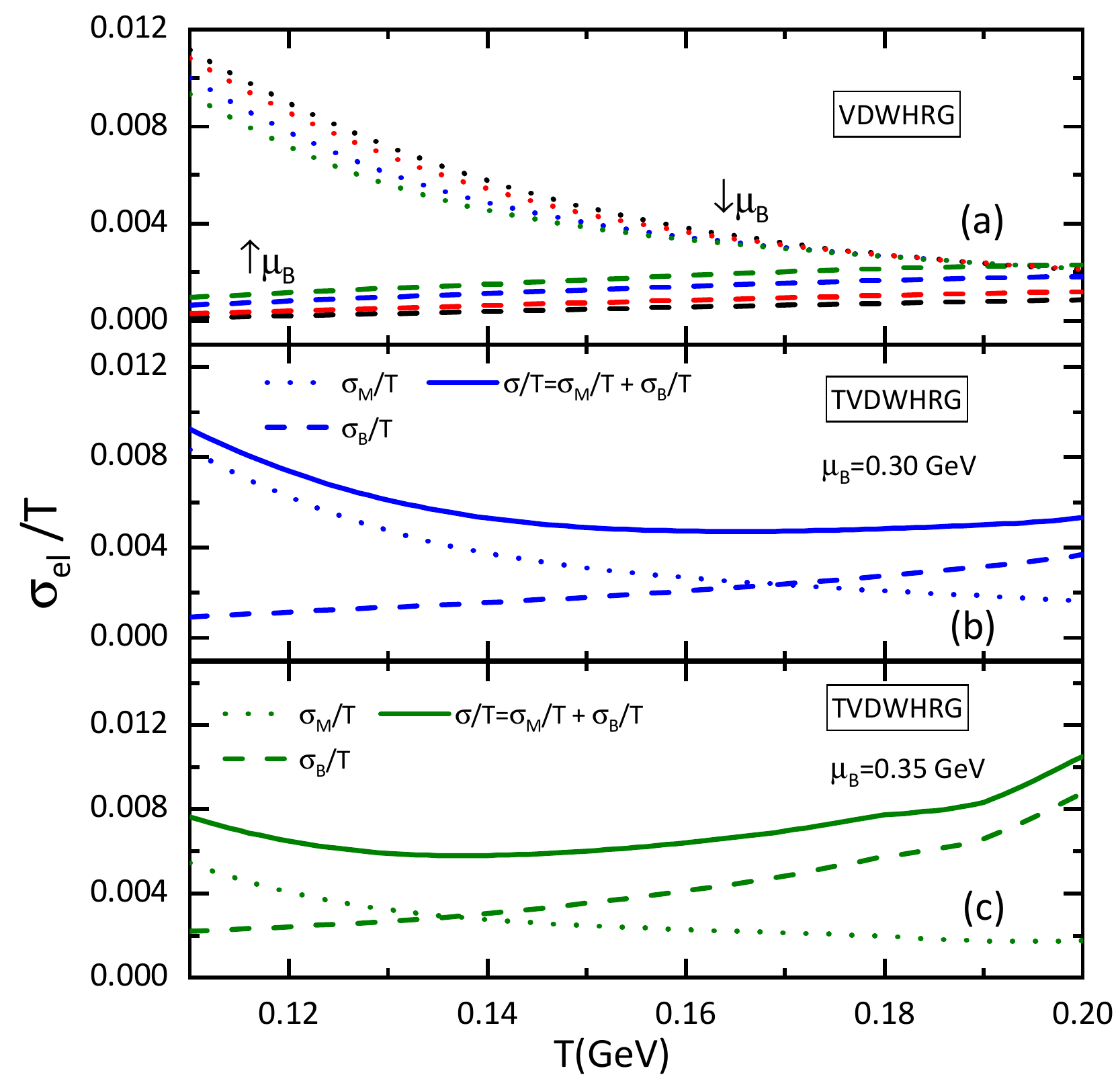}
  	\caption{\label{fig:dec-sigma}(Color online) The temperature dependence of the scaled electrical conductivity for meson (dotted lines) and baryon components (dashed lines), and their total (solid lines) at various $\mu_{B}$.}
  \end{figure}

  In regard to the scaled electrical conductivity $\sigma_{el}/T$ at vanishing $\mu_{B}$, we compare our result in TVDWHRG model (olive-green solid line) with existing estimations, as shown in Fig.~\ref{fig:sigma-vs}.  
  The pink short dotted line represents the result for pion gas in unitarized ChPT via Green-Kubo technique~\cite{sek-cpt-pion1}.
  The blue open triangles show the result obtained from PHSD approach~\cite{electric-PHSD1}, which is a covariant extension to  Boltzmann-Uehling-Uhlenbeck approach~\cite{BUU}  in hadronic sector.  The cyan solid squares  represent the calculation of $\sigma_{el}/T$ for hadronic  gas employing SMASH  using the Green-Kubo formalism~\cite{electric-smash}.
  The orange dotted-dashed line shows the result in EVHRG model using RTA ~\cite{electric and thermal evhrg}.
  The gray open circles show the computation of  $\sigma_{el}/T$  for $\pi$-$K$-$N$ gas in  kinetic theory (KT) using a  CE-like expansion of the distribution function~\cite{electric-kinetic1}.
  The red full circles are the datas from 2+1 flavor anisotropic lattice QCD calculation~\cite{electric-lattice}. 
  The black dashed line is the estimation of $\sigma_{el}/T$ in a conformal Super-Yang Mills (SYM) theory~\cite{SYM}.
  In Ref.~\cite{electric-kubo}, Ghosh {\it et al} provided an estimation of  $\sigma_{el}/T$ for $\pi$-$N$ system  from electromagnetic current-current correlators in the static limits (brown stars). 
  The bright green diamonds show the result of $\sigma_{el}/T$ in NJL model~\cite{sbet-njl}.
  
  From Fig.~\ref{fig:sigma-vs} we notice that  the  variation  of  $\sigma_{el}/T$ with temperature in  NJL model,  KT and SYM theory  is not obvious,  other model estimations  and  our result  indicate that $\sigma_{el}/T$ for $\mu_{B}=0$~GeV significantly decreases at $T=0.1\sim 0.18$~GeV.
 The result of SMASH~\cite{electric-smash} is roughly 3 times larger than ours.  This is mainly attributed to the choices of calculation methodology and cross sections, as well as the lack of elastic collisions of some possible particle pairs (e.g., elastic $\pi^+\pi^+ $ and $\pi^-\pi^- $) in SMASH. 
 The ChPT result~\cite{sek-cpt-pion1} is close to ours at $T<0.14$~GeV, however, with increasing $T$  the ChPT  result is  a factor of 3 larger  than ours. In ChPT the degrees of freedom are only mesons thus we  deduce that the inclusion of  more hadron species (baryons) may reduce the electrical conductivity  of system.
 The results in NJL model~\cite{sbet-njl} and PHSD model~\cite{electric-PHSD1} are much larger than ours.
  This great deviation  may be  due to the fact that  the elementary  degrees of freedom in the NJL model and PHSD model are (anti-)quarks instead of hadrons. 
 The significant numerical  difference between KT result~\cite{electric-kinetic1} and ours mainly arises from the uncertainties in realistic cross sections and the difference choices in hadron spectrum.
The result of Ghosh {\it et at}~\cite{electric-kubo} is a factor of 2 smaller than ours, which is mainly due to the differences in the inputs of  medium constituents and the relaxation times. The reason of  numerical difference between the estimation of $\sigma_{el}/T$  in EVHRG model~\cite{electric and thermal evhrg}  between ours is similar to what we have discussed earlier about $\eta/s$. It is worth noting that at high $T$ our result is  close to  the Lattice QCD data though our model contains no quark-gluon degrees of freedom.

 Fig.~\ref{fig:sigma} displays the variation of the scaled electrical conductivity  $\sigma_{el}/T$ with respect to
  temperature  at 
 $\mu_{B}=0.1$, $0.2$, $0.3$ and $0.35$ GeV in all considered HRG models.   The $T$ and $\mu_{B}$ dependence of total electrical conductivity is basically coming from the number density and the relaxation time in Eq.~(\ref{eq:sigma}). 
 The number density is more dominating than the relaxation time in determining the $T$ and $\mu_{B}$ dependence of electrical conductivity  for baryonic contribution.
 However, for mesonic contribution, the variation of electrical conductivity in IHRG model with $T$ and $\mu_{B}$ is  primarily governed by the relaxation time rather than the number density. This arises from the mathematical analysis of electrical conductivities of mesons and baryons.
   As we can be seen from Fig.~\ref{fig:sigma}, total $\sigma_{el}/T$ in IHRG model decreases as $T$ increases for  $\mu_{B}=0.1$~GeV. One can understand this behavior as follows: Firstly, the numerical strength of  $\sigma_{el}/T$ in IHRG model mainly comes from the contribution of  mesons. At a given $\mu_{B}$ the contribution of mesons to total $\sigma_{el}/T$, $\sigma_{M}/T$, decreases due to the decrease of relaxation time via scattering with more hard spheres at high $T$. Secondly, the contribution of baryons to total $\sigma_{el}/T$, $\sigma_{B}/T$, increases as $T$ grows although the value of $\sigma_{B}/T$ is very small compared to that of $\sigma_{M}/T$. Thus, after adding  mesonic and baryonic contributions to total $\sigma_{el}/T$, the qualitative  behavior of total $\sigma_{el}/T$ in IHRG model is still dominated by  mesons (pions).
  With the increase of $\mu_{B}$, the baryonic concentration increases and pions scatter with more baryons, leading to  a reduction in the relaxation time of mesons. As a result, $\sigma_{M}/T$ decreases with increasing $\mu_{B}$.  Although $\sigma_{B}/T$ increases with growing $\mu_{B}$, the  increment in $\sigma_{B}/T$ can not win over the reduction in $\sigma_{M}/T$. Hence total $\sigma_{el}/T$ in IHRG model decreases with increasing $\mu_{B}$, as shown in Fig.~\ref{fig:sigma}($a$). 
  Similar to shear viscosity, we also discuss the effects of in-medium hadron masses and VDW interactions on total $\sigma_{el}/T$. As we can  see from Fig.~\ref{fig:sigma}($a$),  at $\mu_{B}=0.2$~GeV  the value of total $\sigma_{el}/T$  is relatively smaller in THRG model than in IHRG model, which is mainly due to  that the reduction in  the relaxation time of mesons within THRG model although   $\sigma_{B}/T$ within THRG model have a slight cancellation effect to the decrease of   $\sigma_{M}/T$. 
  We notice that in  Fig.~\ref{fig:sigma}($a$) at $\mu_{B}=0.1$~GeV or zero $\mu_{B}$ the effect of thermal hadron masses on total $\sigma_{el}/T$ is negligible due to the small in-medium modification of masses, whereas, with the further increase of $\mu_{B}$, the negative impact of thermal hadron masses on total $\sigma_{el}/T$ becomes stronger. So at high  $\mu_{B}$ the effect of in-medium hadron masses on $\sigma_{el}/T$ is significant and  non-ignorable. Nonetheless, the THRG model does not change the qualitative behavior of   $\sigma_{el}/T$.  Hence we can deduce that   total $\sigma_{el}/T$  in IHRG and THRG models is still quantitatively and qualitatively dominated by mesonic contribution, i.e. $\sigma_{M}/T$.

   Next we consider the effect of VDW interactions  on total $\sigma_{el}/T$. In Fig.~\ref{fig:sigma}($b$), total $\sigma_{el}/T$ for $\mu_{B}=0.1$~GeV is significantly enhanced at high $T$ in VDWHRG model  compared to  IHRG model.  
   The reasons are as follows: Firstly, the increase in the relaxation time of mesons due to the inclusion of VDW interactions  makes an enhancement in $\sigma_{M}/T$  at high $T$. Secondly, compared to  IHRG model,  VDWHRG model leads to an  improvement (reduction) in the relaxation time (the number density) of baryons at high $T$.  And the strong rise in the relaxation time of baryons within VDWHRG model makes $\sigma_{B}/T$  has a large enhancement  at high $T$ after dominating over the decrease of the number density of baryons.
Nevertheless,  total $\sigma_{el}/T$  is still decreasing over the entire temperature domain   in  VDWHRG model similar to that in  IHRG model.
 The dependence of $\sigma_{el}/T$ on $\mu_{B}$ in VDWHRG model is non-monotonous, in stark contrast to that in IHRG model, as shown in Fig.~\ref{fig:sigma} ($b$).  More exactly, as $\mu_{B}$ grows  total $\sigma_{el}/T$ in VDWHRG model first decreases at low $T$ then increases at high $T$. In order to better understand this non-trivial behavior, the temperature dependencies of  $\sigma_{B}/T$ and  $\sigma_{M}/T$  within VDWHRG model at various $\mu_{B}$ are plotted in Fig.~\ref{fig:dec-sigma}($a$). 
  At high $T$,  $\sigma_{B}/T$ in VDWHRG model is comparable with  $\sigma_{M}/T$ and the increase of $\sigma_{B}/T$ is enough to compensate the inconspicuous decrease of $\sigma_{M}/T$  with the increase in $\mu_{B}$. Thus at high temperature, the  variation of total $\sigma_{el}/T$ with $\mu_{B}$ is dominated by $\sigma_{B}/T$, as shown in Fig.~\ref{fig:dec-sigma}($a$).
  We also study the mix effects of thermal hadron masses  and VDW interactions on total $\sigma_{el}/T$ at various $\mu_{B}$. In Fig.~\ref{fig:sigma}($b$), we observe that the variation of total $\sigma_{el}/T$ with  $\mu_{B}$ in TVDWHRG model is analogous to  that in VDWHRG model.  It is worth noting that  $\sigma_{el}/T$  in  TVDWHRG model at $\mu_{B}=0.3$ and $0.35$~GeV  shows  a broad hollow with a minimum,   which is  qualitatively similar to the result in  Ref.~\cite{electric-kubo}, where  $\sigma_{el}/T$ for $\pi$-$N$ system is calculated  at $\mu_{N}=0.4$, 0.5 and 0.6~GeV.
  Similarly, the results in PHSD model\cite{electric-PHSD2} and NJL model~\cite{sbet-njl}  show that $\sigma_{el}/T$ at $\mu_{B}=0$~GeV decreases in hadronic temperature  region  but increases in  partonic temperature region  and the minimum of $\sigma_{el}/T$ around the critical temperature.
   This non-monotonous behavior of  $\sigma_{el}/T$  is because   the value of  $\sigma_{B}/T$ for $\mu_{B}=0.3$ and $0.35$~GeV  in TVDWHRG model   significantly  overshoots the value of $\sigma_{M}/T$  at high $T$, as shown in Fig.~\ref{fig:dec-sigma}($b,c$). Therefore, we conclude that at high $\mu_{B}$ the positive effect of the VDW interactions on electrical conductivity will be further improved by the inclusion of thermal hadron masses, even if the thermal mass effect itself leads to a  reduction in electrical conductivity.

  \begin{figure}
	\includegraphics[width=2.5in,height=2.5in]{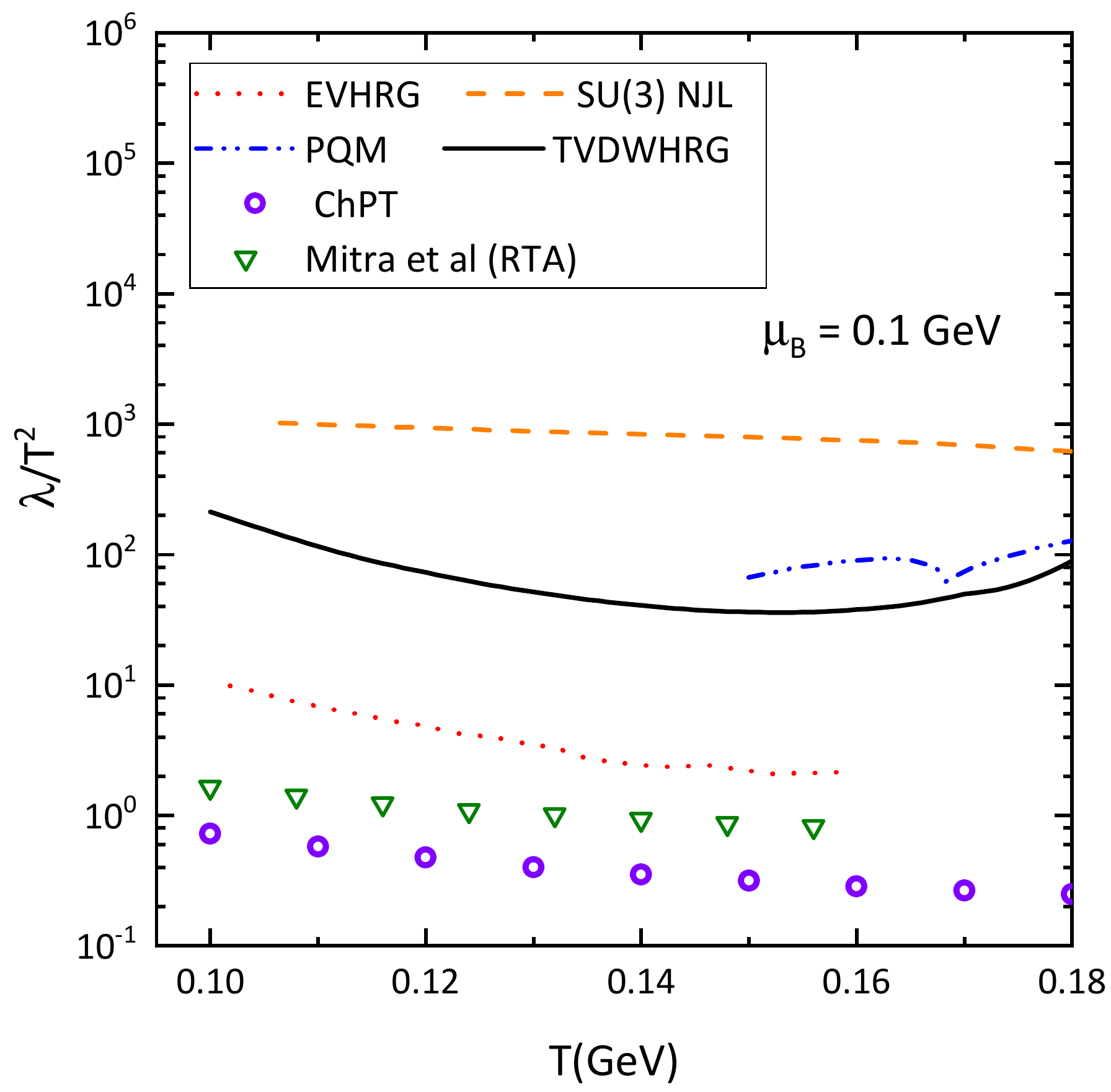}
	\caption{\label{fig:heat-vs}(Color online) Comparison of several calculations for the scaled thermal conductivity $\lambda/T^2$ at $\mu_{B}=0.1$~GeV, see text for more explanations.}
\end{figure}
\begin{figure*}
	\includegraphics[width=5.5in,height=2.5in]{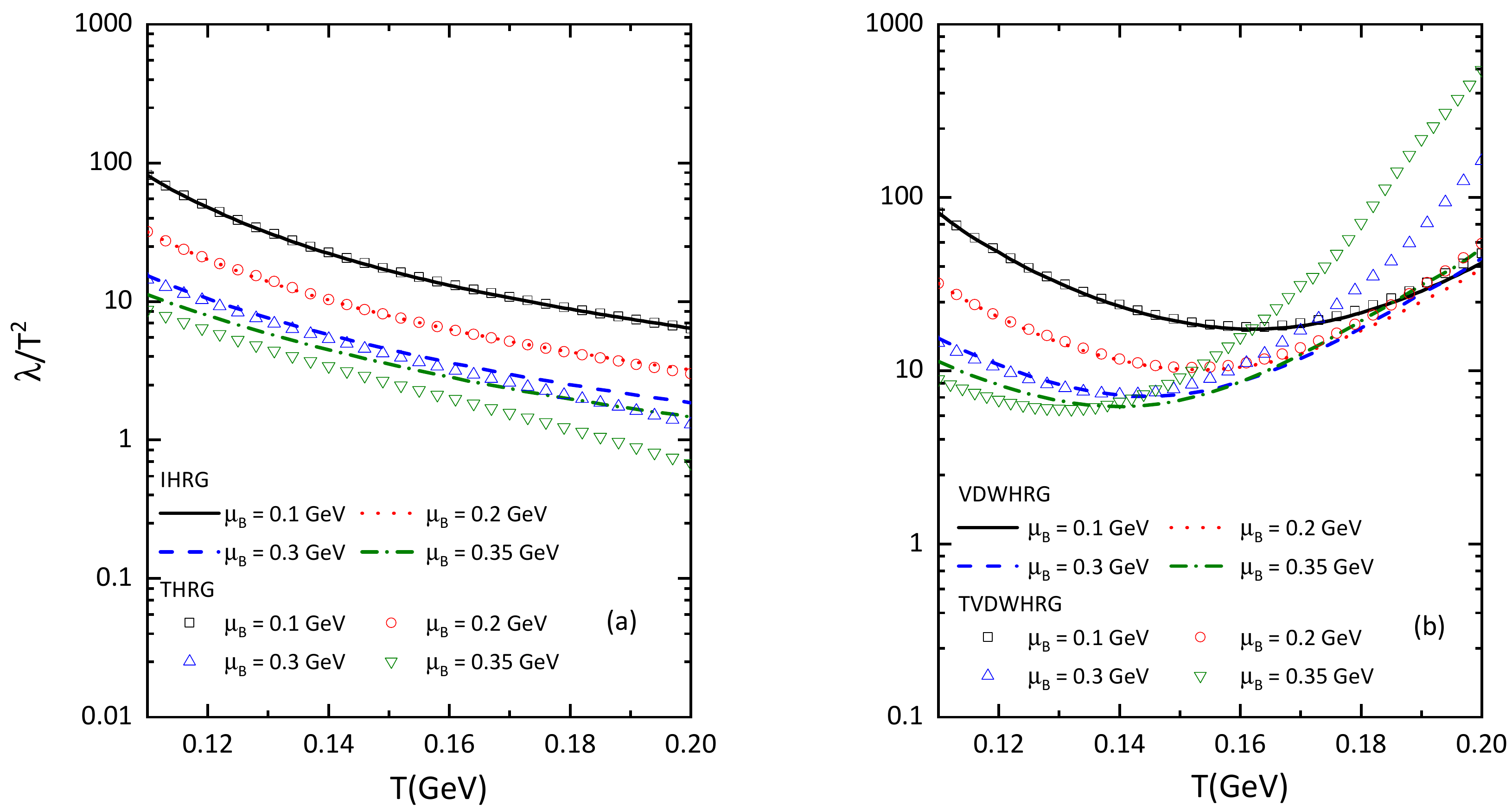}
	\caption{\label{fig:heat}(Color online) Same as Fig.~\ref{fig:Eta} for  the	scaled thermal conductivity $\lambda/T^{2}$.}
\end{figure*}

Fig.~\ref{fig:heat-vs} displays the temperature dependence of the scaled thermal conductivity $\lambda/T^{2}$ within TVDWHRG model at $\mu_{B}=0.1$~GeV (black solid line). We remind the reader that  in a baryon-free ($n_{B}=0$) hadronic system, there is no thermal conduction which is related to the relative flow of energy  and baryon number, hence thermal conductivity vanishes. But for pure pion gas with conserved number,  thermal conductivity can be  non-zero  at vanishing  $\mu_{B}$~\cite{RTA1}.
We  also compare our result with the results of some earlier works.  The orange dashed line and blue double-dotted-dashed line   correspond to the estimation of $\lambda/T^{2}$ at $\mu_{B}=0.1$~GeV in SU(3) NJL model~\cite{sbet-njl} and in  SU(2) Polyakov Quark Meson (PQM) model~\cite{sbe-PQM model}, respectively.   The red dotted line represents the result in EVHRG model~\cite{electric and thermal evhrg}. The green triangles represent the estimation of $\lambda/T^{2}$ by Mitra {\it et al} for pion gas using RTA~\cite{thermal-kinetic-pion}. The purple open circles show the result for pion gas in unitarized ChPT using Green-Kubo formalism~\cite{sek-cpt-pion1}. We notice our result is more or less  in qualitative similar with these existing results.
Whereas, the  calculations of various models have significantly different orders of magnitude. The estimation of $\lambda/T^{2}$ by Mitra {\it et al}~\cite{thermal-kinetic-pion} and the ChPT result~\cite{sek-cpt-pion1} are far less than ours, since total $\lambda/T^{2}$ in pion gas is only coming from   $\pi$-$\pi$ elastic scatterings. The numerical difference between the result in EVHRG model and our result may again be attributed to the fact that in Ref.~\cite{electric and thermal evhrg} the repulsive interactions are related to  all hadrons rather than only baryon-baryon pairs and antibaryon-antibaryon pairs.  In addition, the results of $\lambda/T^{2}$ in NJL~\cite{sbet-njl} and PQM models~\cite{sbe-PQM model} are larger than ours since  the elementary degrees of freedom in NJL model (PQM model) are quarks (quarks and light mesons) whereas the  degrees of freedom in HRG models  are  hadrons.

The temperature dependence of  $\lambda/T^{2}$ for  $\mu_{B}=0.1$,~0.2,~0.3,~0.35~GeV within all considered HRG models is plotted in Fig.~\ref{fig:heat}. At a given $\mu_{B}$ the monotonically decreasing behavior of  $\lambda/T^{2}$ in IHRG model 
is  in a large part qualitatively determined by the heat function $w/n_{B}$ as shown in  Eq.~(\ref{eq:heat}).  Furthermore, at a given temperature $\lambda/T^{2}$ decreases as $\mu_{B}$ increases within IHRG model.
 This mainly arises from that the baryon density $n_{B}$ increases by the significant amount with increasing $\mu_{B}$, although  the enthalpy density $w$ also increases as $\mu_{B}$ grows, this effect is small.
 In Fig.~\ref{fig:heat} ($a$) for high $\mu_{B}$ ($viz$, $\mu_{B}=0.3$ and 0.35~GeV),  $\lambda/T^{2}$ in THRG model is reduced quantitatively  compared to that in IHRG model. 
 This is because  although the values of  both $w$ and $n_{B}$ have an enhancement by the inclusion of in-medium hadron masses,  the enhancement of $n_{B}$ is so large that  $w/n_{B}$ in THRG model  as a whole has a reduction  compared to that in IHRG model. At small  $\mu_{B}$ ($\mu_{B}=0.1$ and 0.2~GeV) or zero $\mu_{B}$, $\lambda/T^{2}$ is nearly unaffected by the inclusion of in-medium hadron masses.  This is mainly  due to that  with the consideration of thermal hadron masses the increase in $n_{B}$  is nearly  neutralized by the decrease in $1/w$.  Hence the in-medium hadron masses play an important role in the calculation of $\lambda/T^{2}$ especially at high $\mu_{B}$. 

We observe that the qualitative variation of $\lambda/T^{2}$ with $T$ and $\mu_{B}$ in THRG model is akin to that in IHRG model, as shown in Fig.~\ref{fig:heat}($a$). However, when we consider the effect of VDW interactions   on $\lambda/T^{2}$, its behavior becomes unusual. In Fig.~\ref{fig:heat} ($b$),  $\lambda/T^{2}$ for $\mu_{B}=0.1$~GeV in VDWHRG model first decreases, reaches a minimum, then increases with increasing temperature, which is not observed in EVHRG model~\cite{electric and thermal evhrg}. And the minimum of $\lambda/T^{2}$ for $\mu_{B}=0.1$~GeV around $T=0.16$~GeV.  Similarly, $\lambda/T^2$  in PQM model~\cite{sbe-PQM model} and  NJL model~\cite{thermal-NJL} for $\mu_{B}=0.1$~GeV also shows a  non-monotonous behavior with a minimum near the critical temperature. This valley structure of $\lambda/T^{2}$ in VDWHRG model may be explained as follows:  At low $T$ the hadronic system is dominated by light  mesons  whose contributions to  $\lambda/T^{2}$ are nearly not affected by VDW interactions. Thus at low $T$, the $T$ and $\mu_{B}$  dependence of $\lambda/T^{2}$ in VDWHRG model mimics that in IHRG model.  With  increasing  $T$ the baryonic states increases, the VDW interactions leads to  a reduction in both $w$ and $n_{B}$, however the reduction of $n_{B}$ is so prominent that makes  $\lambda/T^{2}$ be an increasing function of $T$  in high $T$ domain. 
In short, the $\mu_{B}$ dependence of $\lambda/T^{2}$ in VDWHRG model is still  analogous to that in THRG and IHRG models. We also notice that the minimum  of  $\lambda/T^{2}$ in VDWHRG model shifts to lower temperature as $\mu_{B}$ increases.  Furthermore, the effect of the VDW interactions on  $\lambda/T^{2}$ is more pronounced  by the inclusion of thermal hadron masses at high $\mu_{B}$ ($\mu_{B}=0.3$ and 0.35~GeV), even though the effect of thermal  masses itself can result in  a numerical decrease of  $\lambda/T^{2}$. Thus  for $\mu_{B}=0.3$ and 0.35~GeV   $\lambda/T^{2}$ in TVDWHRG model increases faster at high $T$   compared to that in VDWHRG model and  the value of  $\lambda/T^{2}$ at $\mu_{B}=0.35$~GeV even overshoots   the value of  $\lambda/T^{2}$ at $\mu_{B}=0.3$~GeV, which can be shown in Fig.~\ref{fig:heat}($b$).

\begin{figure}
	\includegraphics[width=2.9in,height=1.7in]{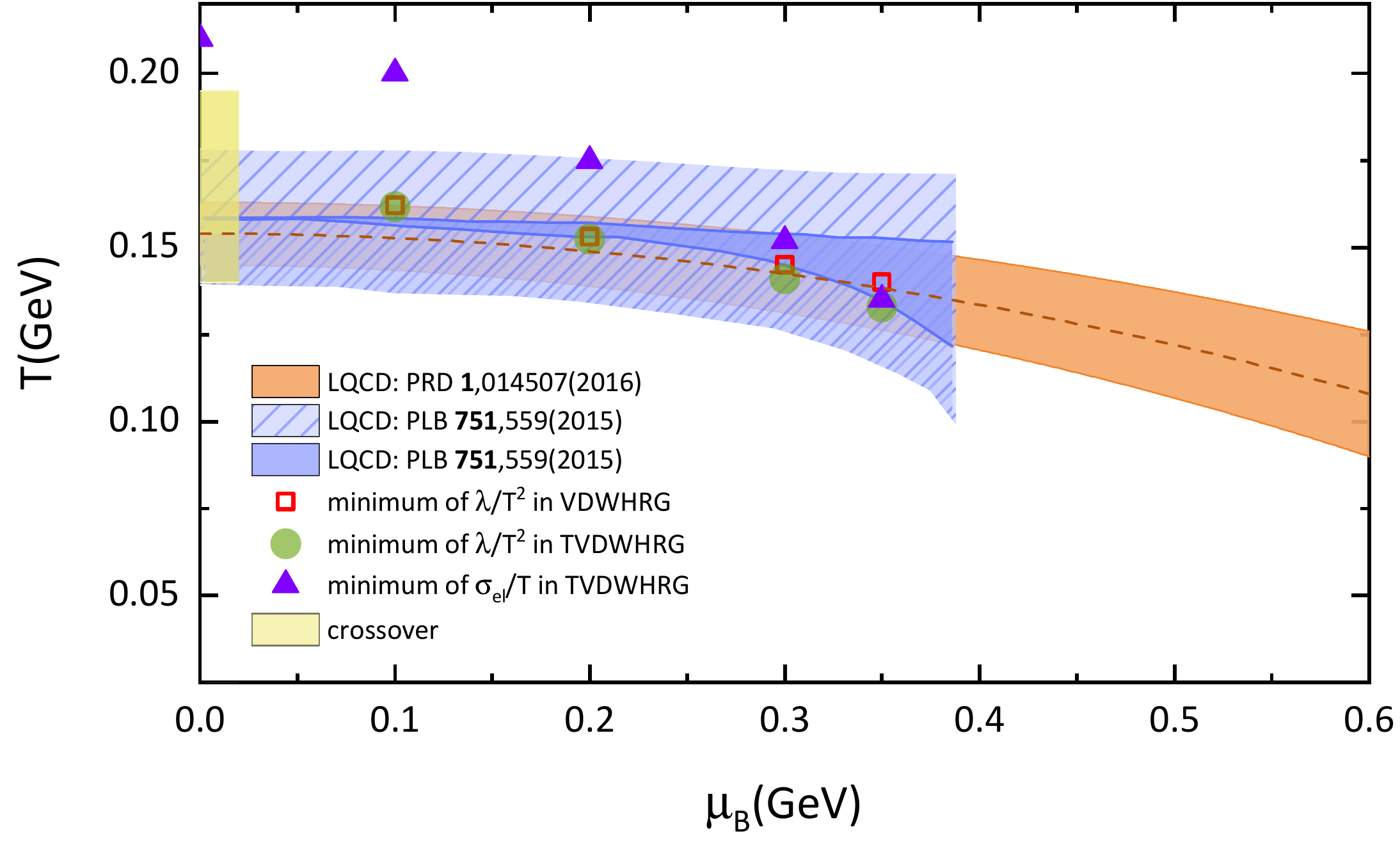}
	\caption{\label{fig:phase}(Color online) The minima for $\sigma_{el}/T$ in VDWHRG model (purple  solid traingles) and minima for $\lambda/T^2$ in VDWHRG model (orange open squares) and TVDWHRG model (green solid circles) in the $T-\mu_{B}$ plane.  
	The phase diagram obtained by analytic continuation of Lattice QCD simulations from imaginary to real $\mu_{B}$~\cite{Bellwied:2015rza,Cea:2015cya}. The blue dashed band and orange solid band indicate the width of the phase transition. The blue solid band is the critical line from Ref.~\cite{Bellwied:2015rza}. The widening around 0.3~GeV is coming from the uncertainity of the curvature and from the contribution of higher order. The orange dashed line shows the transition line from Ref.~\cite{Cea:2015cya}
	The yellow band is expected crossover region ($T=0.14\sim0.19$~GeV) for $\mu_{B}=0~\mathrm{GeV}$ in Ref.~\cite{Aoki:2006br}. }
\end{figure}

Fig.~\ref{fig:phase} shows the minima of $\sigma_{el}/T$ for TVDWHRG model  and the minima of  $\lambda/T^2$  for VDWHRG and TVDWHRG models in the $T-\mu_{B}$ plane.
We notice that these minima are  phenomenologically located  inside or slightly deviate the phase transition region obtained from Lattice QCD simulations~\cite{Bellwied:2015rza,Cea:2015cya}. For $\lambda/T^2$, the minima in VDWHRG and TVDWHRG models are very close to the critical transition lines.
Based on previous results in  Refs~\cite{sbet-njl,electric-PHSD2,thermal-NJL,sbt-PQM model} where the minimum of  $\sigma_{el}/T$  ($\lambda/T^2$) is near the critical temperature at $\mu_{B}=0$~GeV ($\mu_{B}=0.1$~GeV),
   we expect the scaled electrical and thermal conductivities in TVDWHRG model at different $\mu_{B}$ exhibit a minimum near the QGP-hadron phase transition region making it a crucial signature of the phase transition.  Up to our knowledge,  there are no results of the scaled conductivities based on Lattice QCD calculations and the effective QCD  models at different non-zero $\mu_{B}$, so whether the minimun  is really a sign of phase transition still needs to be verified in the future.

\vspace*{.6cm}
\section{CONCLUSION}
\label{sec:conclusions}

 In this work we investigate the thermodynamics and transport coefficients with the thermal van der Waals hadron resonance gas (TVDWHRG) model, which is the extension of VDWHRG model by including the  effect of temperature $T$ and baryon chemical potential $\mu_{B}$ dependent hadron masses.
 In TVDWHRG model thermal hadron masses are obtained  by 2+1 
 flavor Polyakov linear sigma model combined with the scaling rule of hadron masses. 
 We estimate the thermodynamics, such as  the pressure, the  energy density, the entropy density and the square of sound velocity in TVDWHRG model and compare them with the Lattice QCD data.
 It has been shown that  at $T\sim 0.16-0.195$~GeV the thermodynamics for $\mu_{B}=0$ GeV in  TVDWHRG model give an improved agreement with the available Lattice QCD data  compared to that in VDWHRG model. 
 And with the increase of $\mu_{B}$, the thermodynamics, e.g. the pressure,  have a sizeble improvement in magnitude  due to the inclusion of thermal hadron masses.
   
 We also investigate the scaled transport coefficients,  such as shear  viscosity to the entropy density ratio $\eta/s$, the scaled electrical conductivity  $\sigma_{el}/T$, and the scaled thermal conductivity $\lambda/T^{2}$ of hadronic matter in all considered HRG models, by using the quasi-particle kinetic theory under relaxation time approximation. 
 From the qualitative  and quantitative perspectives,
 taking into account the effects of VDW interactions and thermal hadron masses, the scaled transport coefficients are modified considerably.
 When we only consider the effect of $T$ and $\mu_{B}$ dependent   hadron masses, compared to  IHRG model case, the values of all the  scaled transport coefficients  for  fixed $\mu_{B}$  are relatively suppressed in THRG model even though $\eta$ itself is enhanced in THRG model. 
 Though  the suppression of the scaled transport coefficients due to thermal mass effect is relatively weak at small $\mu_{B}$ ($viz$, $\mu_{B}=0$,~0.1,~0.2~GeV),  with the increase of $\mu_{B}$ its effect becomes more pronounced. Nonetheless, the general behaviors of transport coefficients in THRG model and  in IHRG model are similar qualitatively. 
 
  However, compared to IHRG model,  VDWHRG model  leads to a qualitatively and  quantitatively different behavior of  the scaled transport coefficients. On the one hand, the  VDW interactions between (anti)baryons give a significant enhancement of the scaled transport coefficients at high $T$ and even change the dependence of $\lambda/T^{2}$ on temperature. 
  On the other hand, as $\mu_{B}$ grows  $\eta/s$ and $\sigma_{el}/T$ in VDWHRG or TVDWHRG model decrease at low $T$ whereas increase  at high $T$.
  Furthermore, the effect of VDW interactions on the scaled transport coefficients for $\mu_{B}>0.2$ GeV is strengthened further at high $T$  by the inclusion of in-medium hadron masses  though  thermal hadron masses itself have a negative effect on the scaled transport coefficients. 
  The minimum of $\sigma_{el}/T$ in TVDWHRG  and the minimum of $\lambda/T^{2}$ in TVDWHRG or VDWHRG model may be related to phase transition, which needs to be verified based on the research from different effective models.
 It is noted that we have made some simple assumptions in the present TVDWHRG model, and there are a lot of space for further improvement (e.g. the approaches we obtain all in-medium hadron masses could be improved;  The VDW parameters may vary with $\mu_{B}$; The
  quantitative changes of constituent quark masses in various effective QCD models may be modified, etc). But anyway,  we expect the improved TVDWHRG model does not break down the existing  qualitative behaviors for the scaled transport coefficients in the present TVDWHRG model.

\vspace*{.6cm}
{\it Acknowledgments:}  This research is supported in part by the NSFC of China with Project No. 11935007.

\end{document}